\documentclass{aastex6}

\usepackage{graphicx}
\usepackage{graphics}
\usepackage[figuresleft]{rotating}
\usepackage{caption3}
\usepackage{amsmath}
\usepackage{mathrsfs}

\begin{document}


\title{Wavelet Denoising of Radio Observations of Rotating Radio Transients (RRATs): Improved Timing Parameters for Eight RRATs}



\author{M. Jiang}
\affil{Lane Department of Computer Science and Electrical Engineering \\
 West Virginia University \\
 Morgantown, WV 26506}

\author{B.-Y. Cui\altaffilmark{1}}
\affil{Department of Physics and Astronomy \\
 West Virginia University \\
 Morgantown, WV 26506}

\author{N.A. Schmid\altaffilmark{1}}
\affil{Lane Department of Computer Science and Electrical Engineering \\
 West Virginia University \\
 Morgantown, WV 26506}

\author{M.A. McLaughlin\altaffilmark{1}}
\affil{Department of Physics and Astronomy \\
 West Virginia University \\
 Morgantown, WV 26506}

\and

\author{Z.-C. Cao}
\affil{Lane Department of Computer Science and Electrical Engineering \\
 West Virginia University \\
 Morgantown, WV 26506}


\altaffiltext{1}{Center for Gravitational Waves and Cosmology, West Virginia University,  Morgantown, WV 26505}

\begin{abstract}

Rotating radio transients (RRATs) are sporadically emitting pulsars detectable only through searches for single pulses. While over 100 RRATs have been detected, only a small fraction (roughly 20\%) have phase-connected timing solutions, which are critical for determining how they relate to other neutron star populations. Detecting more pulses in order to achieve solutions is a key to understanding their physical nature. Astronomical signals collected by radio telescopes contain noise from many sources, making the detection of weak pulses difficult. Applying a denoising method to raw time series prior to performing a single-pulse search typically leads to a more accurate estimation of their times of arrival (TOAs). Taking into account some features of RRAT pulses and noise, we present a denoising method based on wavelet data analysis, an image-processing technique. Assuming that the spin period of an RRAT is known, we estimate the frequency spectrum components contributing to the composition of RRAT pulses. This allows us to suppress the noise, which contributes to other frequencies. We apply the wavelet denoising method including selective wavelet reconstruction and wavelet shrinkage to the de-dispersed time series of eight RRATs with existing timing solutions. The signal-to-noise ratio (S/N) of most pulses are improved after wavelet denoising. Compared to the conventional approach, we measure $12\%$ to $69\%$ more TOAs for the eight RRATs. The new timing solutions for the eight RRATs show $16\%$ to $90\%$ smaller estimation error of most parameters. Thus, we conclude that wavelet analysis is an effective tool for denoising RRATs signal.

\end{abstract}




\section{Introduction}
\label{sec:intro}

Rotating Radio Transients (RRATs) are sporadic pulsars that were detected through a search for single pulses and are not detectable through their time-averaged emission. Their profiles are typically a few milliseconds in duration, with intervals between detected pulses ranging from minutes to hours \citep{mll+06}.  They were first discovered through single-pulse search reprocessing of the Parkes Multibeam Pulsar Survey data  \citep{man01a,kel+09,km11} and currently over 100 such objects are known\footnote{See http://astro.phys.wvu.edu/rratalog}. It is clear that RRATs represent one extreme end of the pulsar intermittency spectrum, with some objects detected as RRATs appearing as normal pulsars when observed with a more sensitive telescope or at a different frequency. However, it is not yet clear what physical or environmental properties determine the emission properties of neutron stars.  The discovery of more RRATs, followed by long-term observations resulting in phase-connected solutions, is essential to answer this question.

For most pulsars, in order to calculate times of arrival (TOAs) the time series data are folded \citep{lar96} at the identified period to amplify the signal-to-noise ratio (S/N) of the pulse and to calculate a composite profile. However, the sporadic emission from RRATs means that they are usually not detectable through folding. Therefore, it is necessary to calculate the TOA of each individual pulse to determine timing solutions. However, astronomical signals collected by radio telescopes contain many types of noise including radiometer noise, sky noise, and radio frequency interference (RFI). Even in  remote areas, where most radio telescopes are located, terrestrial sources of RFI can have a significant impact on our ability to detect single pulses,  and in the most extreme cases, terrestrial RFI is so strong that it makes the signals completely undetectable. In order to conduct a successful search for single pulses in the time domain, it is desirable to decrease this background noise in the data before applying a single-pulse search procedure. Applying an efficient denoising method to data containing RRAT signals may lead to an increased number of detected single pulses and improved accuracy of TOAs and timing solutions.

A conventional denoising method was proposed by \citet{cm03}, which presented a framework for processing radio data to detect single pulses. A boxcar filter (matched filter) was utilized to detect pulse candidates in de-dispersed time series, which were then plotted for manual inspection. This conventional method was employed among others by \citet{dcm+09}, \citet{kle+10}, \citet{bnm12} and \citet{cbm+17} for single-pulse searches. 

In this work, we describe a wavelet-based denoising algorithm that may be applied to de-dispersed time series data containing RRAT pulses. We compare the pulsar fitting results based on the proposed method with the results in \citet{cbm+17} which employed a conventional denoising method. The method builds on the assumption that for pulses collected by a radio telescope, the noise contributes to the entire frequency spectrum, while the RRAT pulses mostly contribute to lower frequency bands. Here, the frequency is not a radio frequency or a spin frequency. The frequency mentioned throughout this paper, without any other specification, is the frequency of the Fourier spectrum of the time series. Assuming that the spin period of an RRAT is known, we estimate the range of frequencies contributing to a description of an RRAT signal and use it to develop an effective denoising approach, which leads to an improved timing solution compared to the conventional method. 

Wavelets are functions localized in time and space \citep{dau92}. Similar to Fourier Transforms (FTs), wavelet transforms are linear transforms used to approximate (decompose) any unknown function (signal). Wavelet transforms (WTs) can be used for data compression, signal analysis, data transformation and other applications. In the last decade wavelet analysis had a great impact on mathematics in the fields of approximation theory and harmonic analysis \citep{cc04}, as well as on engineering in the fields of signal processing, image processing, and computer graphics \citep{mal99,cs05}. Different from Fourier analysis, WTs involve functions described by two parameters, delay and scale, leading to comprehensive joint time- and frequency-domain analyses. This feature of wavelets provides us with an efficient way to extract RRAT pulses from astronomical data. In spite of our inability to entirely remove noise from the data, we can decrease its magnitude and thus increase the S/N of the RRAT pulses. 

The remainder of the paper is organized as follows. Section \ref{sec:Observations} describes the eight RRATs analyzed in this paper. In Section  \ref{sec:Denoising}, we introduce WTs. Based on the a priori knowledge of the spin period of a RRAT signal, we develop a frequency description of the RRAT and then apply wavelet denosing using selective wavelet reconstruction and wavelet shrinkage to the data containing the RRAT signal. The denoised time series are then subjected to the conventional single-pulse searching. In Section \ref{sec:Analysis}, we discuss metrics that we apply to evaluate the performance of the wavelet-based denoising approach. The results of denoising, timing analyses, and pulse profiles are summarized in Section \ref{sec:Results}. Section \ref{sec:Discussion} analyzes the S/N distribution of RRAT data before and after the application of the wavelet denoising approach to data and the factors for timing solution improvement. Finally, planned future work and conclusions are presented in Section \ref{sec:Conclusion and Future Work}. 

\section{Observations} 
\label{sec:Observations}

Five (J1048$-$5838, J1739$-$2521, J1754$-$3014, J1839$-$0141, and J1848$-$1243) RRATs discussed in this paper were discovered in the process of re-analyzing the 1.4-GHz  Parkes Multibeam Pulsar Survey data \citep{man01a,kel+09}, PSR J1623$-$0841 was discovered through the 2007 GBT 350-MHz drift-scan survey \citep{blr+13} and the remaining two (J0735$-$6302 and J1226$-$3223) were discovered in re-analyses of the 2009 1.7 GHz southern-sky high Galactic latitude survey data \citep{jbo+09,bb09}. The follow-up timing and monitoring observation used the same two telescopes and are reported in \citet{cbm+17}. The eight RRATs have spin periods ranging from $0.4$ to $6.2$ s. The observation frequencies of the follow-up observations cover the range from 350 MHz to 1.4 GHz and are summarized in Table \ref{tab1} .

\begin{table*}[!t]
\centering
\resizebox{\textwidth}{!}{%
\hspace{-3cm}
\begin{tabular}{lcccccccccc}
\tableline\tableline
PSR Name  & Telescope & Data Machine &  Frequency & Bandwidth & Sample Time & Time Span & Number & Burst Rate & Mean Flux Density\\
      & & & (MHz) & (MHz) & ($\mu s$) & (Years) & of Observations & ($\rm (hr^{-1}$) & for a Single Pulse (mJy)\\
\tableline
J0735$-$6302  & Parkes & BPSR & 1400 & 256 & 64.0 & 2.06  & 22  &  62.5  & 0.4\\   
J1048$-$5838  & Parkes & SCAMP & 1400 & 256 & 100.0 & 15.2 & 52  & 6.0 &  2.4\\
J1226$-$3223  & Parkes & BPSR & 1400 & 256 & 64.0 & 1.94  & 19  &  49.3 & 2.8\\
J1623$-$0841  & GBT & GUPPI & 350/820 & 100/200 & 245.76 & 2.43 & 24  & 45.0 & 11 (820 MHz); 84 (350 MHz)\\
J1739$-$2521  & GBT & GUPPI & 820 & 200 & 491.52 & 1.60 & 25  &  38.3 	& 49\\
J1754$-$3014  & GBT & GUPPI & 350/820 & 100/200 & 245.76 & 1.56  & 25  &  120.3	& 21 (820 MHz); 410 (350 MHz)\\
J1839$-$0141  & GBT & GUPPI & 820 & 200 & 491.52 & 2.50  & 38  &  19.5 	& 18\\
J1848$-$1243  & GBT & GUPPI & 820 & 200 & 245.76 & 1.80  & 29  &  34.0 	& 13\\
\tableline
\end{tabular}}
\caption{Observing parameters for the eight RRATs described in this paper. \label{tab1}}
\end{table*}

\section{Wavelet Denoising} 
\label{sec:Denoising}

\begin{figure}[!t]
\centering
\vspace{-0.3cm}
\includegraphics[width=0.8\textwidth]{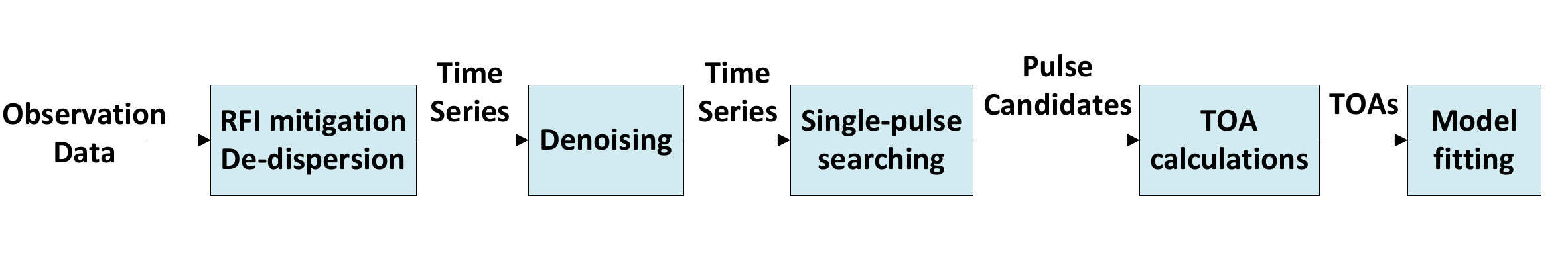} 
\vspace{-0.45cm}
	\caption{Block diagram of the conventional single-pulse search processing applied to a time series containing an RRAT's signal. In this paper, we focus on the last four blocks. }
	\label{fig:data_process_diagram}
\end{figure}

Figure \ref{fig:data_process_diagram} shows a block diagram of the conventional single-pulse processing applied to a time series containing an RRAT signal. In this paper, we focus on the last four blocks in Figure \ref{fig:data_process_diagram}, assuming that the data have been already de-dispersed. We will later demonstrate in Section \ref{sec:Results} that the inclusion of wavelet denoising leads to improved detection of single pulses and allows for more precise timing results compared to those obtained with the conventional approach.

\subsection{Wavelet Transform}  
\label{subsec:Wavelet_Transform}
 
The wavelet denosing approach proposed in this paper includes two steps. Selective wavelet reconstruction, or selecting the proper wavelet decomposition sub-bands for reconstructing the signal, is the first step. The second step is applying wavelet shrinkage (thresholding) to the wavelet coefficients of the sub-bands selected in the first step. A detailed description of both steps is provided in Sections \ref{subsec:selective wavelet reconstruction} and \ref{subsec:Wavelet Shrinkage}, respectively. 

Wavelets are a set of functions localized in time and space. Similar to sine and cosine functions, wavelets are used as basis functions for a linear decomposition of non-stationary or time varying signals. A wavelet is defined as:
\begin{equation} \label{eq:wavelet_define}
\psi _{a,b}(t) =  \frac{1}{\sqrt{a}} \psi \left ( \frac{t-b}{a} \right ),
\end{equation}
where $\psi (\cdot )$ is a known function, $a$ is a time scale parameter, and $b$ is a delay parameter. The parameter $a$ is a positive real number, selected from a set $A$; $b$ is a real number, selected from a set $B$. Note that $\psi _{1,0}(t)$ is known as the mother wavelet, selected to satisfy (see \citet{dau88} for details): 
\begin{equation}\label{eq:wavelet_condition}
\int_{-\infty }^{\infty }\psi (t)dt=0.
\end{equation}
A great multiplicity of mother wavelet functions has been developed in the past two decades. The choice of a mother wavelet function depends on a specific application. 
 
Similar to a Fourier decomposition, which approximates a function by a linear combination of trigonometric functions with different frequencies, wavelets employ the scale parameter $a$ to analyze the frequency content of a signal. Given a mother wavelet and a set of scaled and delayed wavelet functions, the continuous WT \citep{dau90} of a function $f\left ( t \right )\in L^{2} $ at scale $a$ and delay $b$ is given as: 
\begin{equation}\label{eq:wavelet_trans}
W_{\psi }f\left ( b,a \right )= \int_{-\infty }^{\infty }f(t)\cdot \psi^{*} _{a,b}(t)dt ,
\end{equation}
where $W_{\psi }f\left ( b,a \right )$ are wavelet coefficients, and ``$*$'' denotes the operation of complex conjugate. The scale and delay parameters are often set to powers of 2 such that $a = 2_{ }^{-s}$ and $b = k2_{ }^{-s}$, where $k$ and $b$ are integers.   

Because the WT has two parameters--scale and delay -- a signal processed by the WT yields an output signal with two dimensions: frequency and time, respectively. This explains how wavelet analysis combines both the frequency- and time-domain analyses \citep{sn96}. Equation (\ref{eq:wavelet_trans}) shows how WT coefficients are obtained. Here, the collection of all possible pairs $(a,b)$ is denoted by the set $A\times B$, and consequently, the inverse WT \citep{gc99} can be expressed as
\begin{equation}\label{eq:selective_wavelet_recons}
\hat{f}\left ( t \right ) = \frac{1}{C_{\psi }} \iint_{\left (A\times B\right)} \frac{1}{a^{2}} \left [ W_{\psi }f\left (a,b \right ) \right ]\cdot\psi _{a,b}\left ( t \right )\,da\,db,  
\end{equation}
where $\hat{f}\left ( t \right )$ is the selective wavelet reconstruction for $ f\left ( t \right )$. The factor $C_{\psi }$ is a constant depending on the choice of the wavelet and is given by:
\begin{equation}\label{eq:wavelet_constant}
C_{\psi }=\int_{-\infty }^{\infty }\frac{\left | \widehat{\psi }\left ( \omega  \right ) \right |^{2}}{\left | \omega  \right |} d\omega <\infty,
\end{equation}
where the wavelet window $\widehat{\psi}(0)=0$, $``{\ \widehat{•}\ }"$  denotes the operation of FT.

\begin{figure}[!t]
\centering
\vspace{-0.3cm}
\includegraphics[width=9cm,,height=8cm]{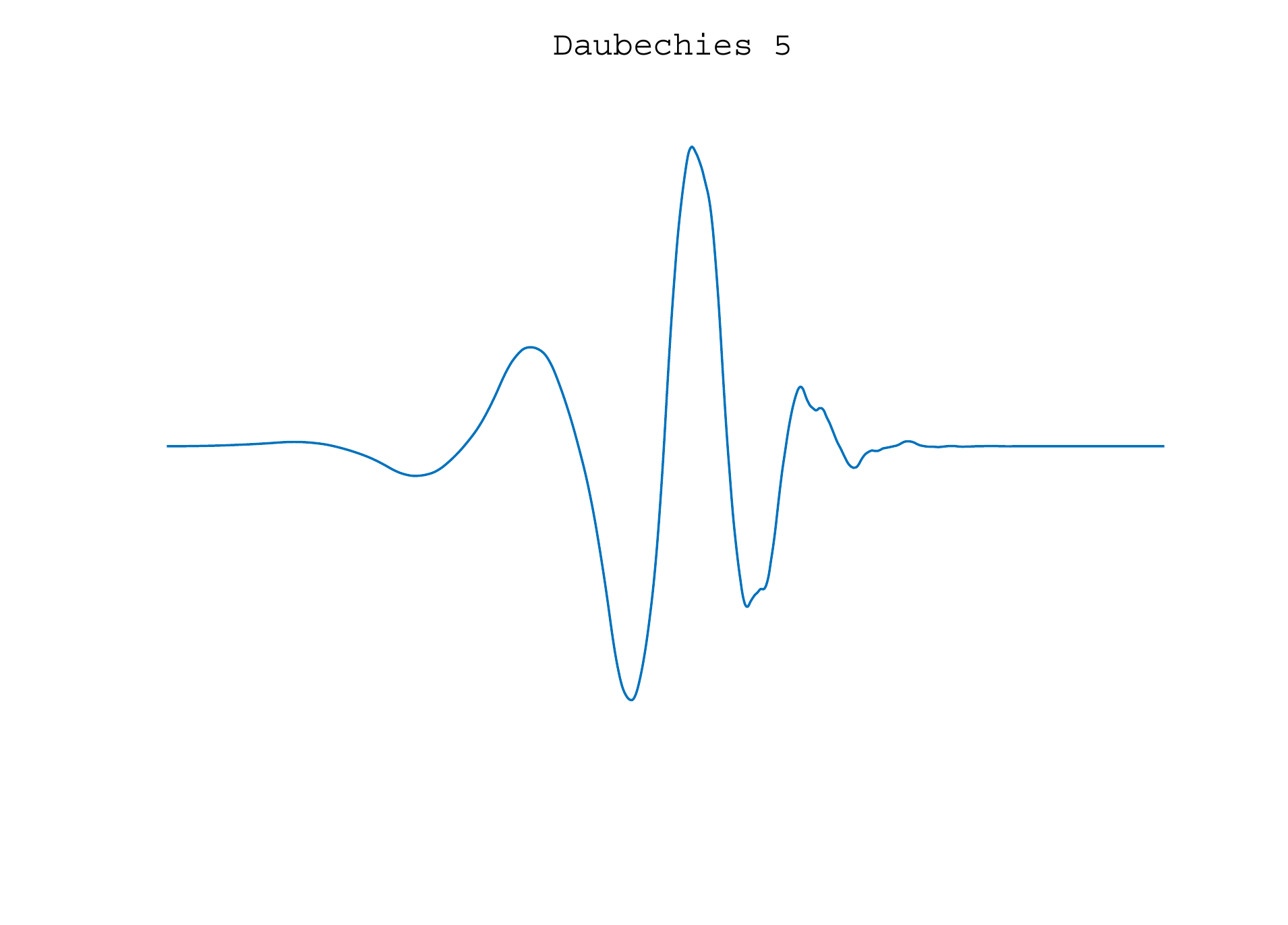} 
\vspace{-2.1cm}
	\caption{The profile of the \textit{Daubechies 5} wavelet. Based on our numerical analysis, a mother wavelet with vanishing moment 5 was found to provide a good representation for a RRAT's pulse. }
	\label{fig:db5}
\end{figure}

When developing a wavelet denoising approach for a particular application, one is expected to (a) select a mother wavelet, (b) summarize selective wavelet reconstruction and (c) specify a wavelet shrinkage (thresholding) approach. Among the most popular mother wavelets are Haar, Daubechies, biorthogonal, Mexican hat, etc. (see \citet{haa10,gm84,dau92,mal99}). They have been utilized for various applications due to their different mathematical properties. \citet{dau92} proposed a series of mother wavelets called \textit{Daubechies n}, where \textit{n} is from 1 to 20, where the integer \textit{n} stands for numbers of vanishing moments. Vanishing moments determine the profile of wavelet. Based on our numerical analysis, a mother wavelet with vanishing moment 5 was found to provide a good representation for an RRATs pulse, and thus we selected \textit{Daubechies 5} as a mother wavelet. Figure \ref{fig:db5} shows the profile of the \textit{Daubechies 5} wavelet. Other wavelet families may be also suitable for representing some RRATs' pulses due to the various single-pulse shapes. The comparison of denoised results of RRATs' data processed  with different wavelet families is beyond the scope of this paper.

\subsection{Selective Wavelet Reconstruction}  
\label{subsec:selective wavelet reconstruction}

After applying the WT with varying values of the scale parameter $a$ to noisy data, we represent the data in terms of ``frequency bands,'' with each band parameterized by a value of $a$. Therefore, if we know the frequency range of an astrophysical pulse of interest, we can extract it from the noisy data by keeping only the appropriate sub-bands and filtering out everything else.

In practice, we usually encounter discrete signals. Therefore, we consider an effective implementation applicable to discrete signals called the Discrete Wavelet Transform (DWT). The DWT uses multiresolution analysis (MRA) and synthesis for signal decomposition and reconstruction \citep{dau90}. MRA allows an efficient implementation of the WT (similar to the FFT for FT). In MRA, a function is viewed at various levels of resolutions, which can be simply understood as the function being represented by wavelets with different scale parameters and delay parameters \citep{gc99,mal89}. 

\begin{figure}[!t]
\centering
\includegraphics[width=0.6\textwidth]{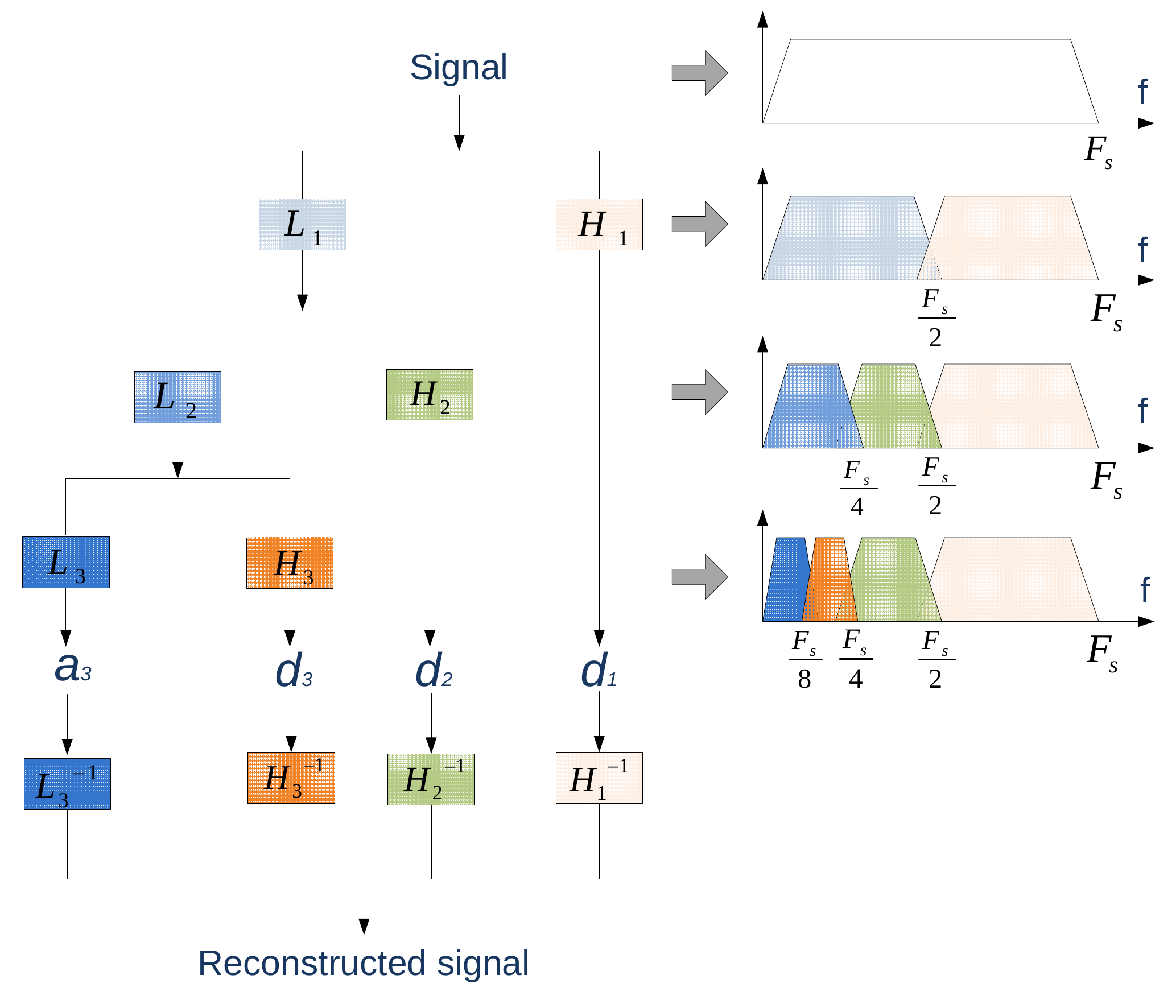} 
\vspace{-0.3cm}
	\caption{The left side of the figure is a discrete wavelet decomposition and reconstruction tree. Here, we chose a three-level tree for our example. The input signal is a time series. $L_n$ and $H_n (n=1,2,3)$ are ``high pass'' and ``low pass'' filters with different cut-off frequencies used to decompose the signal at different scales. Here, $d_n$ is known as the detail wavelet coefficient and $a_n$ is an approximation wavelet coefficient. They are the outcomes of discrete wavelet transformation. $L_{n}^{-1}$ and $H_{n}^{-1}$ are filters used for signal reconstruction (inverse DWT). The right side of the figure shows how the signal is being decomposed by each filter. }
	\label{fig:log_tree}
\end{figure}

Figure \ref{fig:log_tree} shows a discrete wavelet decomposition and reconstruction tree. Here, we selected a three-level tree as an example. The input signal is a time series. $L_n$ and $H_n (n=1,2,3)$ are ``high pass'' and ``low pass'' filters with different cut-off frequencies used to decompose the signal at different scales. In equation (\ref{eq:wavelet_trans}), $\psi^{*} _{a,b}(n)$ is a function used for transforming (decomposing) $f(n)$ at various scales $a$ and different delays $b$. Here, in discrete wavelet decomposition, $L_n$ and $H_n$ are the filters (functions) used for decomposing the signal at various levels of resolution and different delays. They act as $\psi^{*} _{a,b}(t)$ in equation (\ref{eq:wavelet_trans}). Also, $d_n$ is known as the detail wavelet coefficient and $a_n$ is an approximation wavelet coefficient. They are the outcomes of discrete wavelet transformation. $L_{n}^{-1}$ and $H_{n}^{-1}$ are filters used for signal reconstruction (inverse DWT). Different from continuous-time wavelet decomposition, discrete-time wavelet decomposition can be represented by a logarithmic filter tree. 

A single pulse from an RRAT has a shape and duration. Astronomers model the pulse as an intrinsic Gaussian function convolved with an exponential, as the pulse will be scattered due to propagation through the interstellar medium. Due to its shape and duration, a pulse can be described by a narrow range of frequencies. Therefore, after the application of wavelet decomposition, the RRAT signal will be represented by only a few sub-bands in the wavelet decomposition. In order to decide which wavelet sub-bands contribute to an RRAT pulse and thus should be used for reconstruction, we need to first estimate the range of frequencies that contains most of the RRAT signal. 

Consider a time series of length $T$ with samples spaced $t_s$ apart (in our case $t_s$ and $T$ are typically around $100$ $\mu s$ and 30~minutes). If the series contains a few pulses, we can estimate its period $P$. In FT space, the interval between two samples of Fourier spectrum is $1/T.$  The fundamental frequency of the RRAT signal is $f_{min} = 1/P$ and the total number of harmonics of the RRAT signal in the frequency range from $0$ to $1/t_s$ is $ P/t_s$. Because the Fourier spectrum (harmonics) of the RRAT pulses rapidly decrease with increasing frequency, the last (or highest frequency) harmonics are completely dominated by noise. Thus, removing this noise will denoise the RRAT signal. Here, we recommend keeping the $200$ largest frequency components (harmonics) composing of the RRAT signal. This number is selected empirically by using a maximum S/N criterion based on tests of RRATs data. Thus, the main frequency components contributing to the description of RRAT pulses are between $f_{min} = 1/P$ and $f_{max} = 200/P.$ Now we can estimate the decomposition levels in the logarithmic filter tree (see Figure \ref{fig:log_tree}) containing the RRAT signal following the rules:   
\begin{equation} \label{eq:cal_wavelet_select_level}
\begin{aligned}
N{_{max}} = \left \lfloor\log_{2}({F{_{s}}/f{_{min}}})\right \rfloor+1  = \left \lfloor\log_{2} \left( P/t_s\right) \right \rfloor + 1 , \\
N{_{min}} = \left \lfloor\log_{2}({F{_{s}}/f{_{max}}})\right \rfloor+1 = \left \lfloor\log_{2}({P/(200 \times t_s)})\right \rfloor+1 ,
\end{aligned}
\end{equation}      
where $N{_{min}}$ is the lowest wavelet decomposition sub-band (level) retained for reconstruction, $N{_{max}}$ is the highest sub-band (level) retained for reconstruction, and $F{_{s}}$ is the sample frequency of the observation data. 

\begin{figure}[!htb]
\centering
\begin{minipage}[b]{0.32\textwidth}%
\centering \includegraphics[width=5.2cm,height=3.7cm]{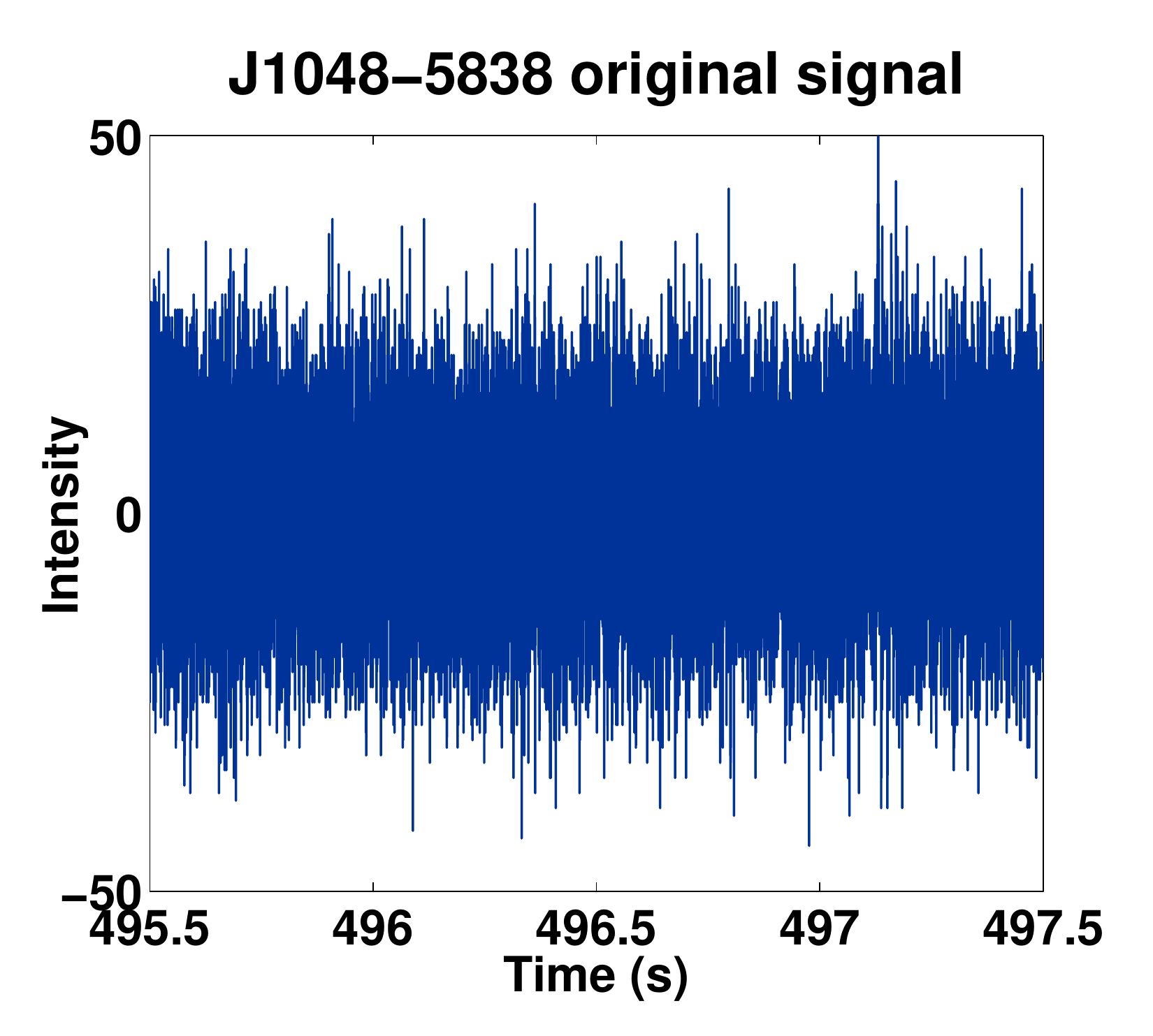}
\end{minipage}%
 \begin{minipage}[b]{0.32\textwidth}%
\centering \includegraphics[width=5.2cm,height=3.7cm]{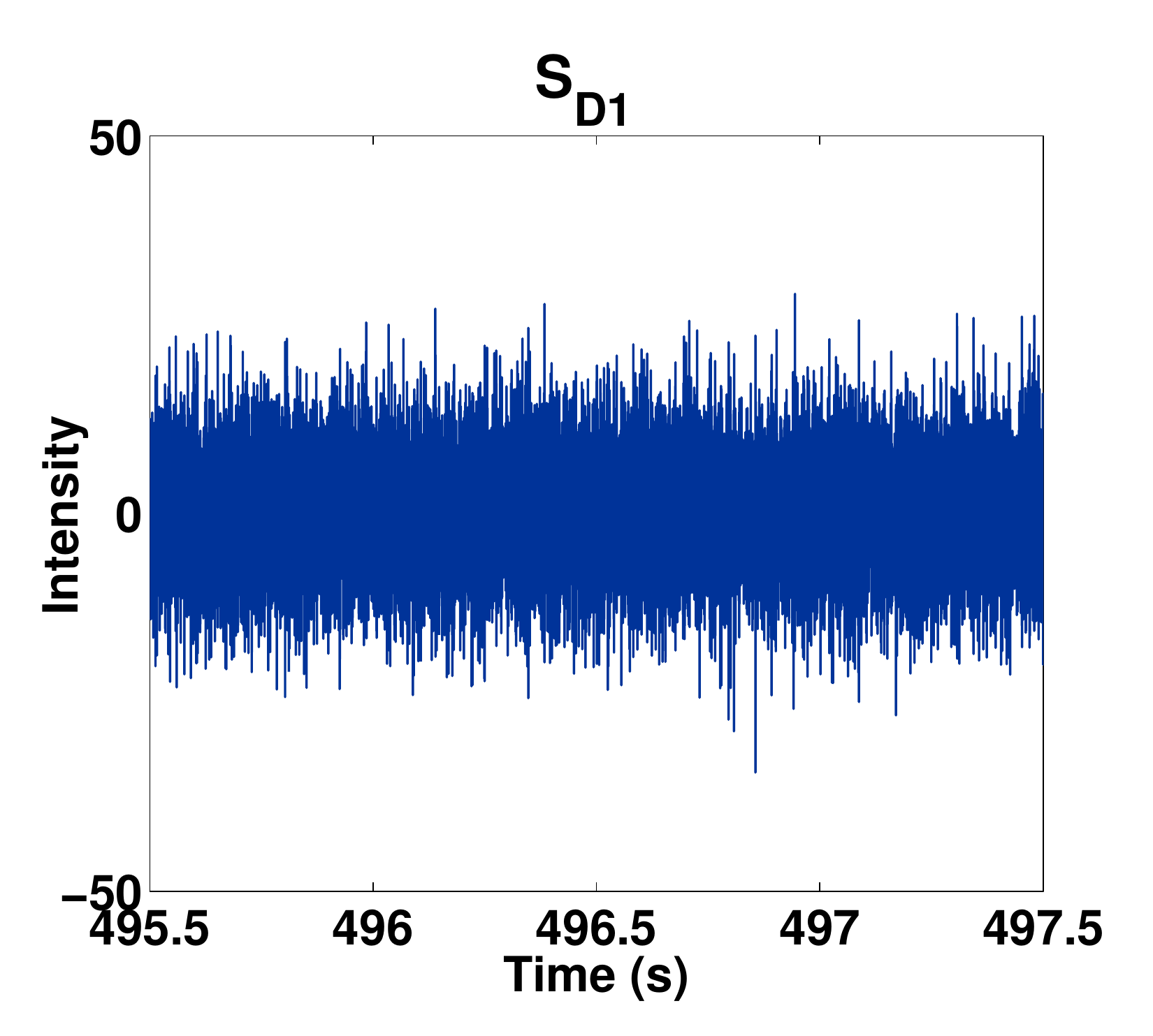}
 \end{minipage}
 \begin{minipage}[b]{0.32\textwidth}%
\centering \includegraphics[width=5.2cm,height=3.7cm]{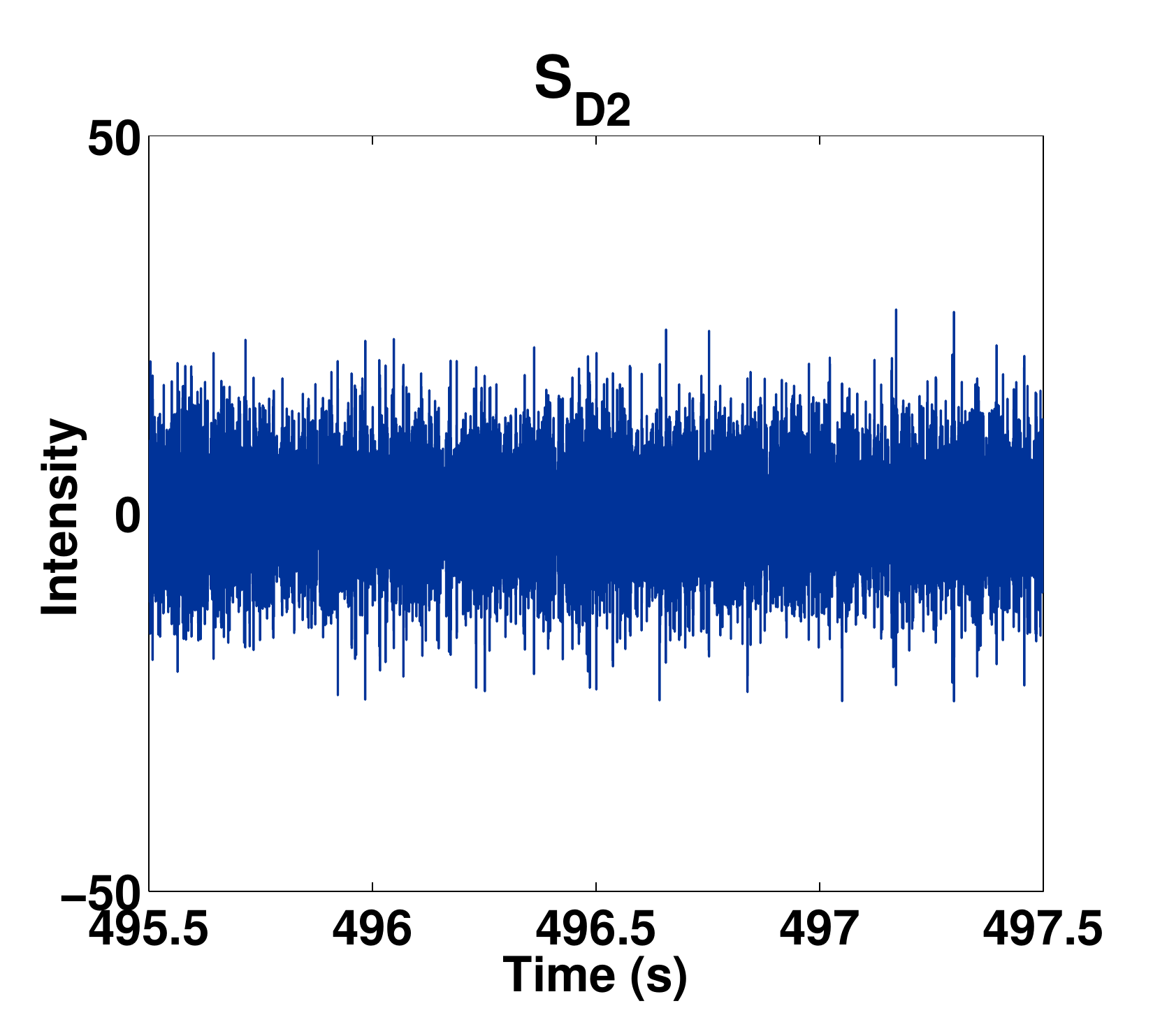}
 \end{minipage}
\vfill
\vspace{-0.15cm}
 \begin{minipage}[b]{0.32\textwidth}%
\centering \includegraphics[width=5.2cm,height=3.7cm]{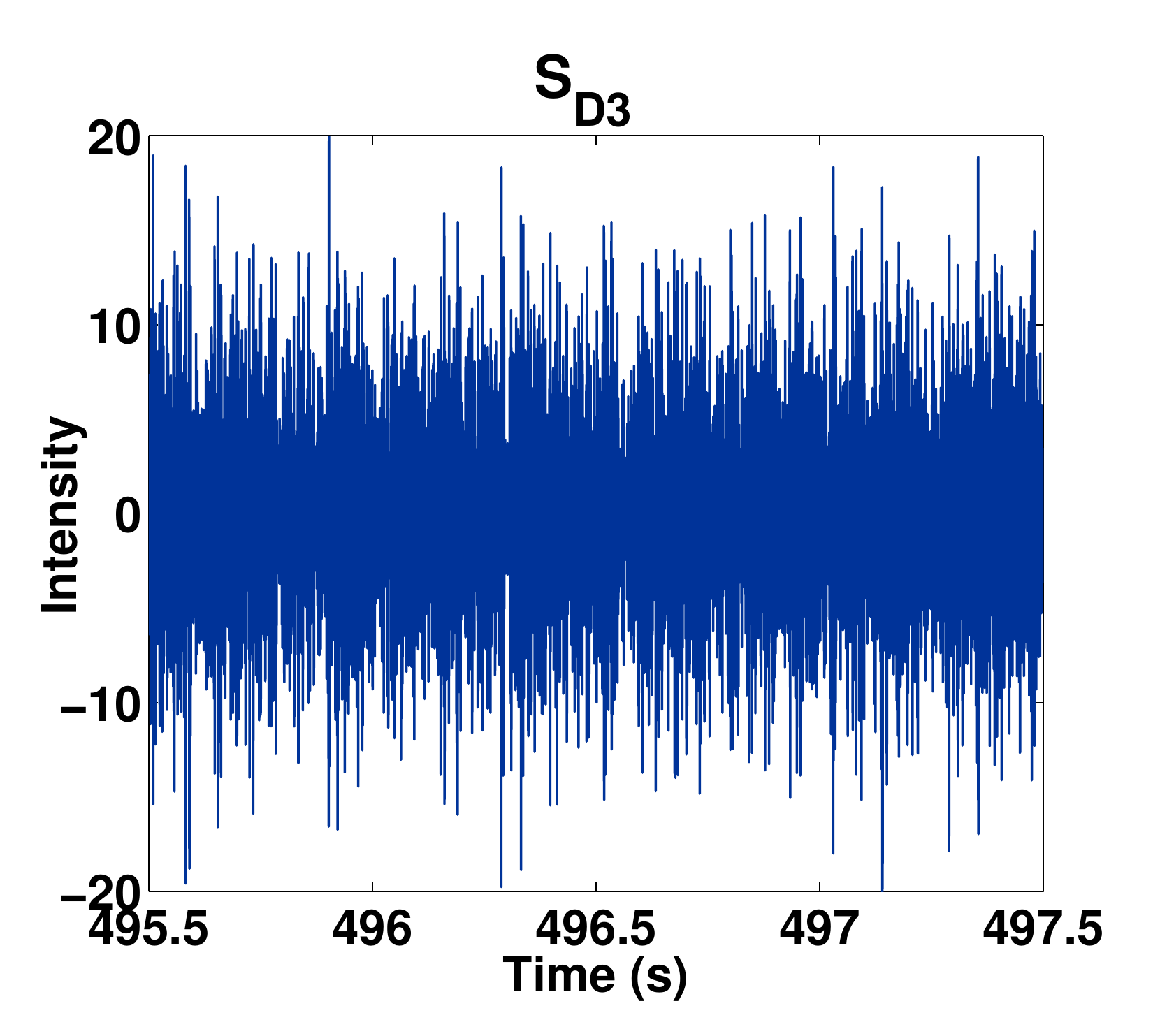}
 \end{minipage}%
 \begin{minipage}[b]{0.32\textwidth}%
\centering \includegraphics[width=5.2cm,height=3.7cm]{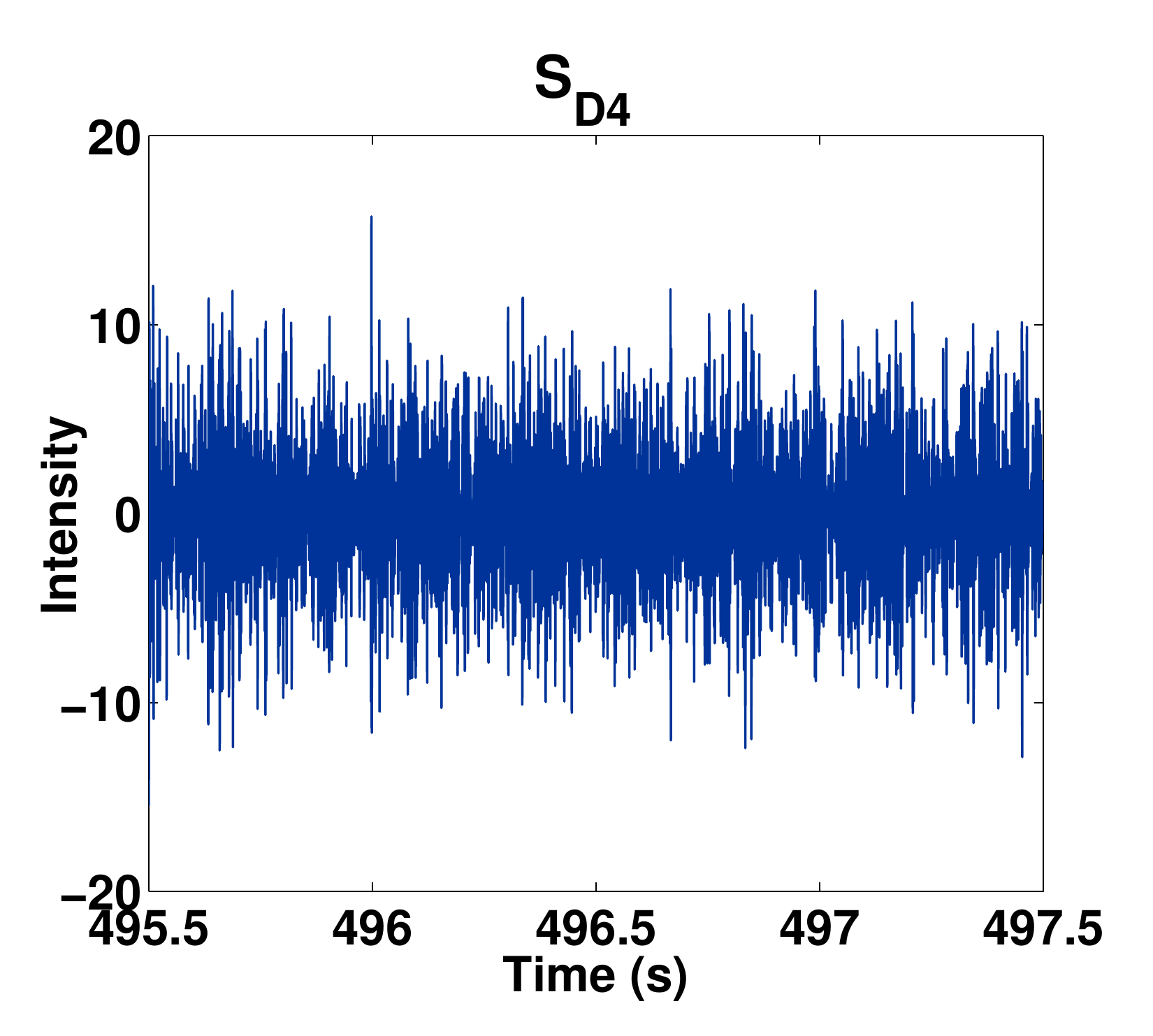}
 \end{minipage}
 \begin{minipage}[b]{0.32\textwidth}%
\centering \includegraphics[width=5.2cm,height=3.7cm]{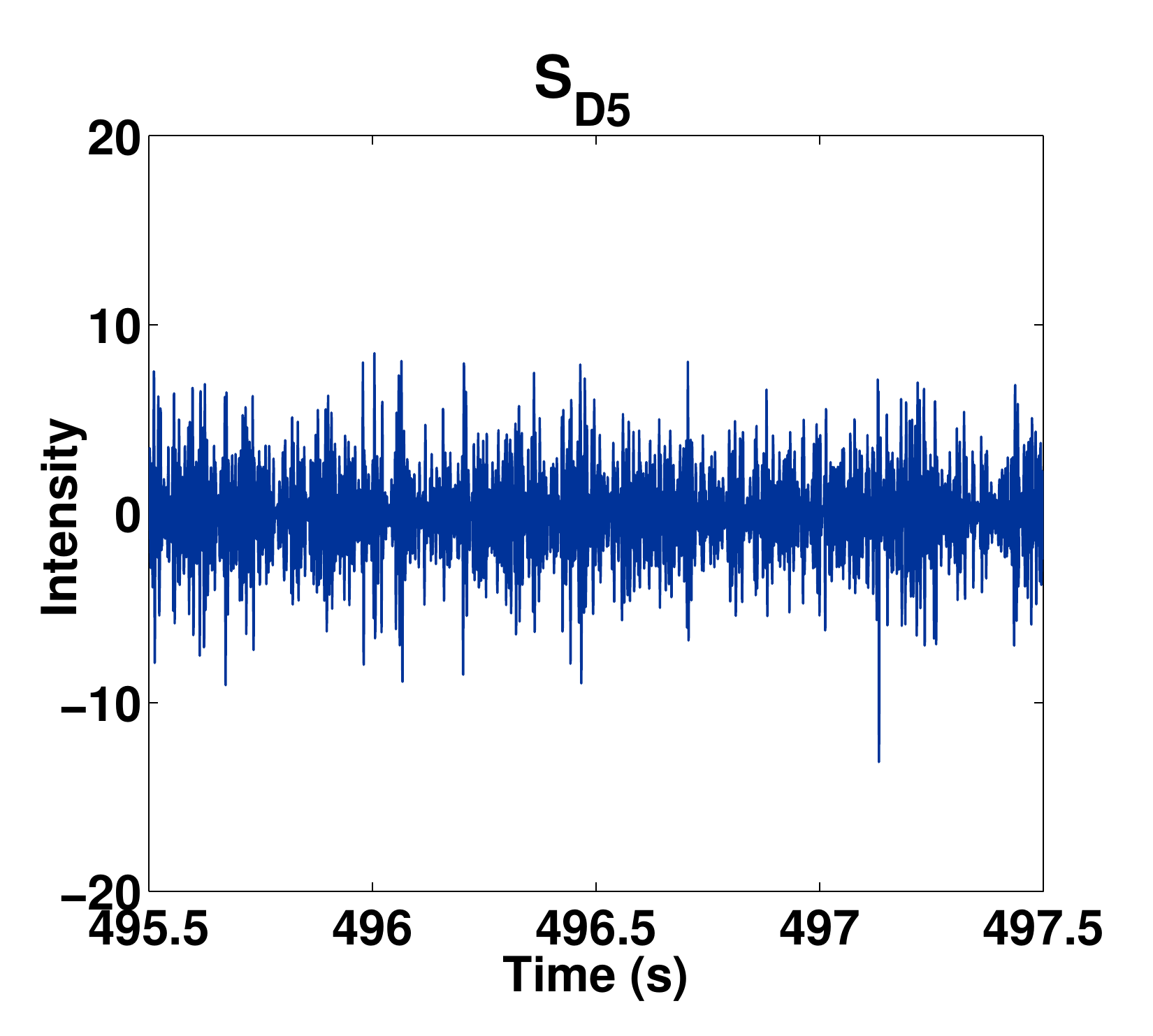}
 \end{minipage}
 \vfill
\vspace{-0.15cm}
 \begin{minipage}[b]{0.32\textwidth}%
\centering \includegraphics[width=5.2cm,height=3.7cm]{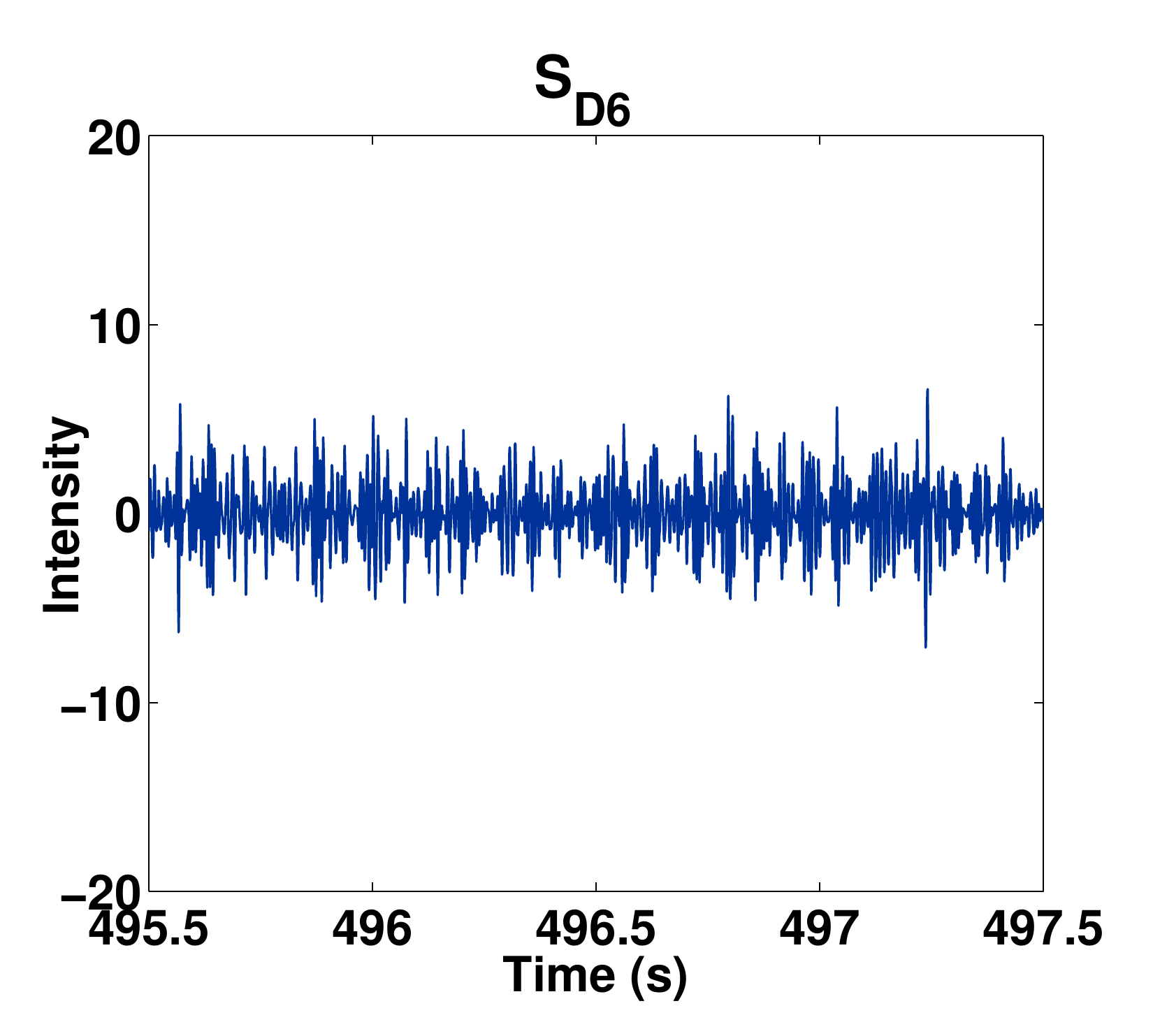}
 \end{minipage}%
 \begin{minipage}[b]{0.32\textwidth}%
\centering \includegraphics[width=5.2cm,height=3.7cm]{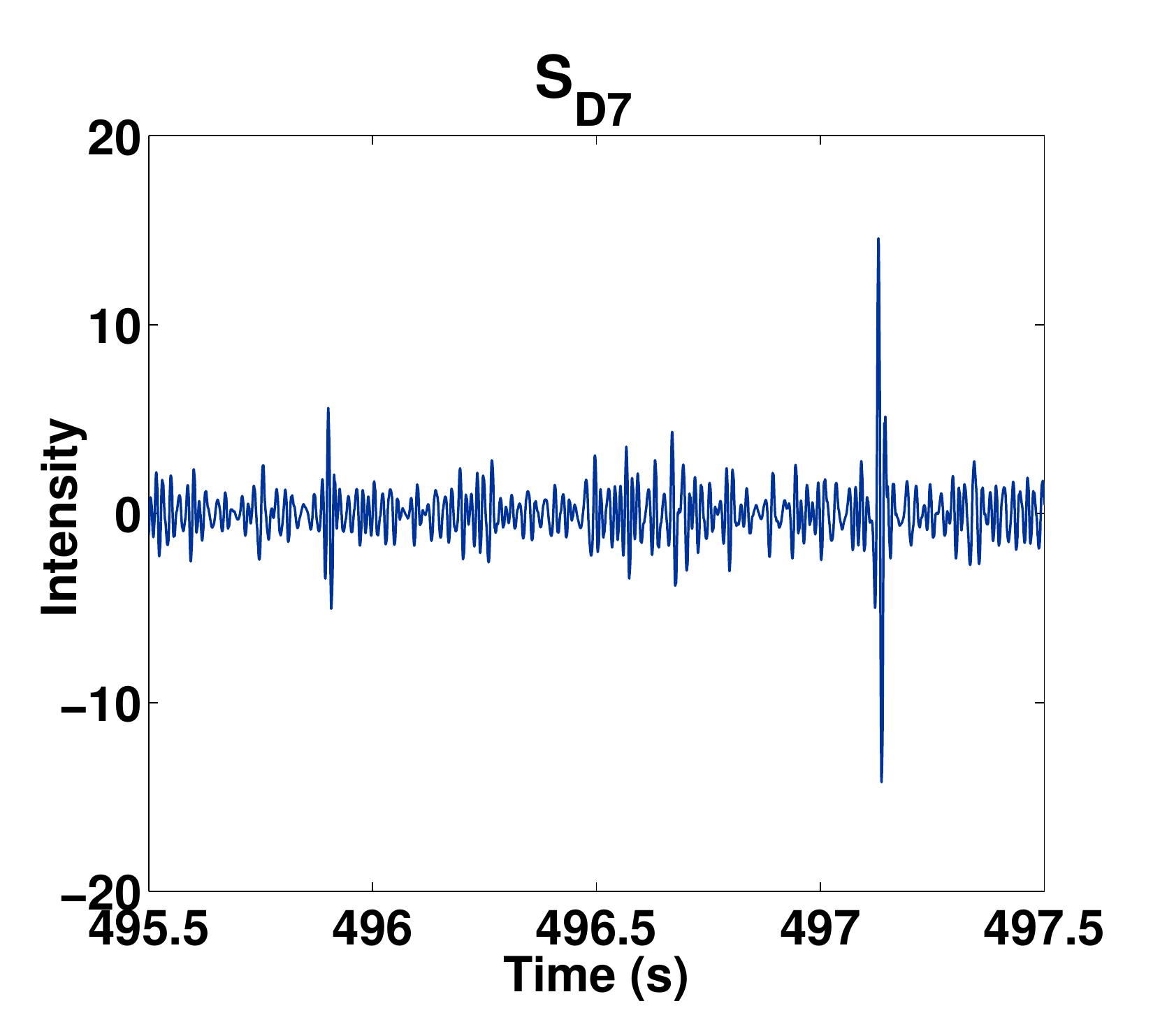}
 \end{minipage}
 \begin{minipage}[b]{0.32\textwidth}%
\centering \includegraphics[width=5.2cm,height=3.7cm]{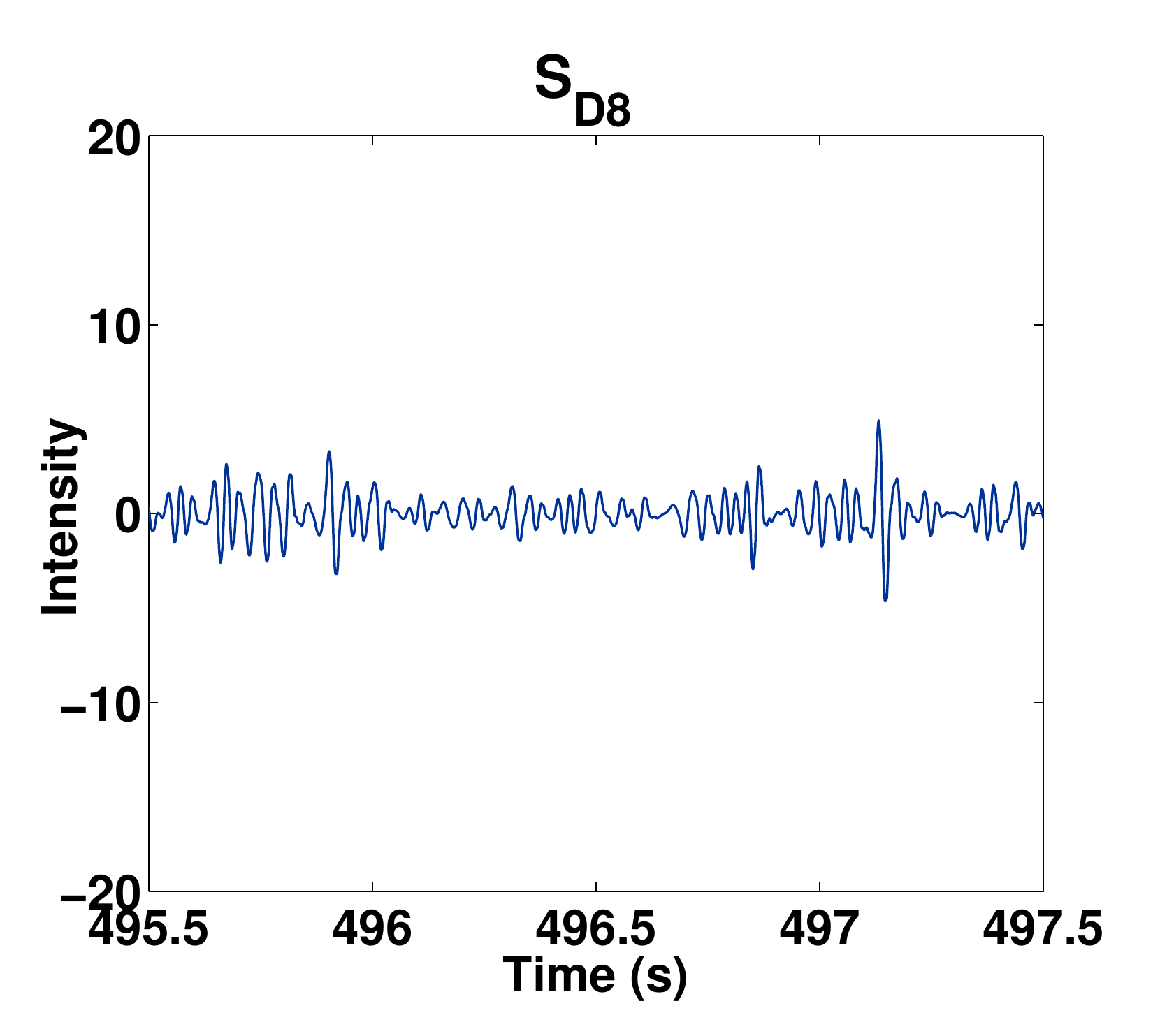}
 \end{minipage}
\vfill
\vspace{-0.15cm}
 \begin{minipage}[b]{0.32\textwidth}%
\centering \includegraphics[width=5.2cm,height=3.7cm]{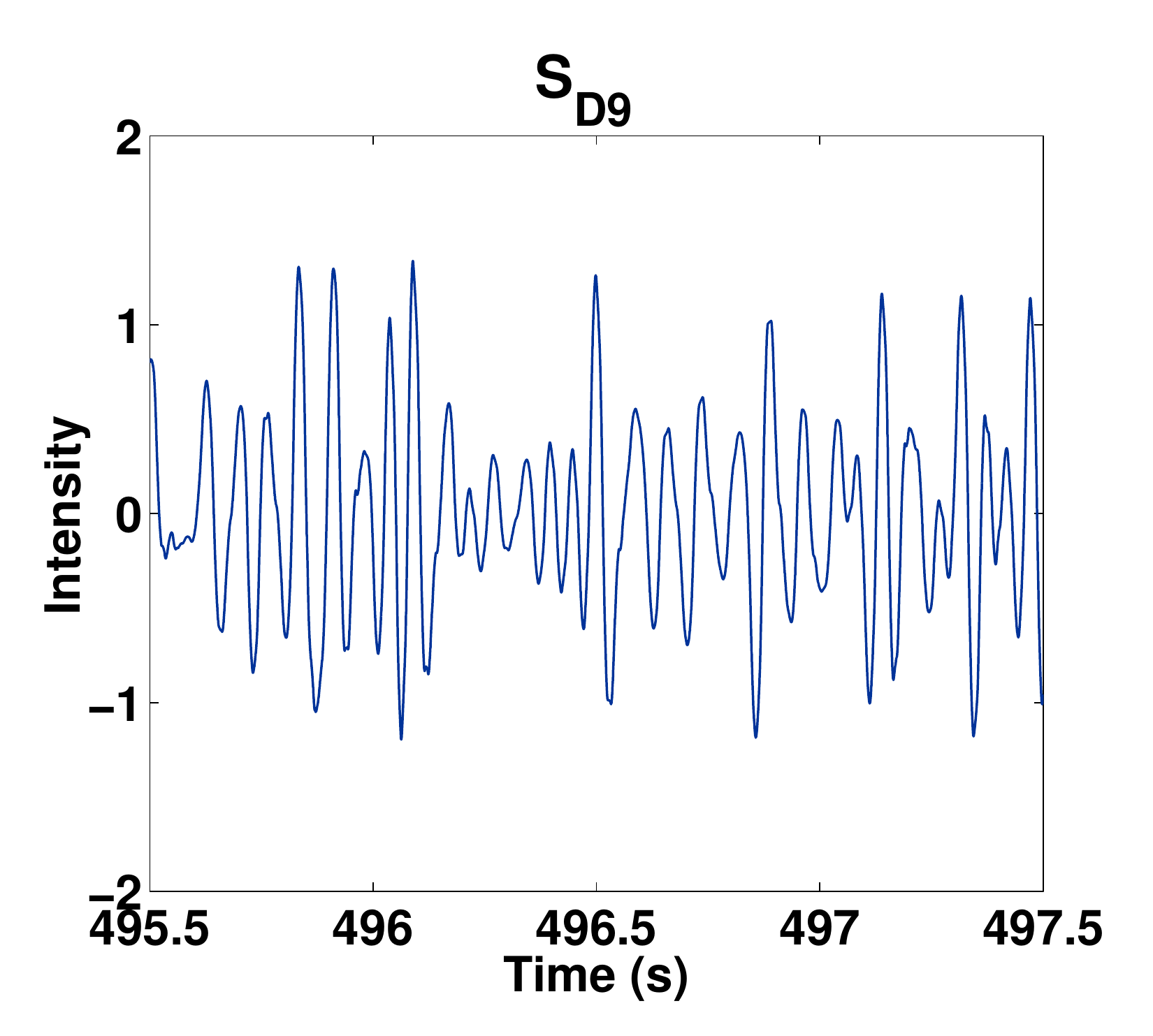}
 \end{minipage}%
 \begin{minipage}[b]{0.32\textwidth}%
\centering \includegraphics[width=5.2cm,height=3.7cm]{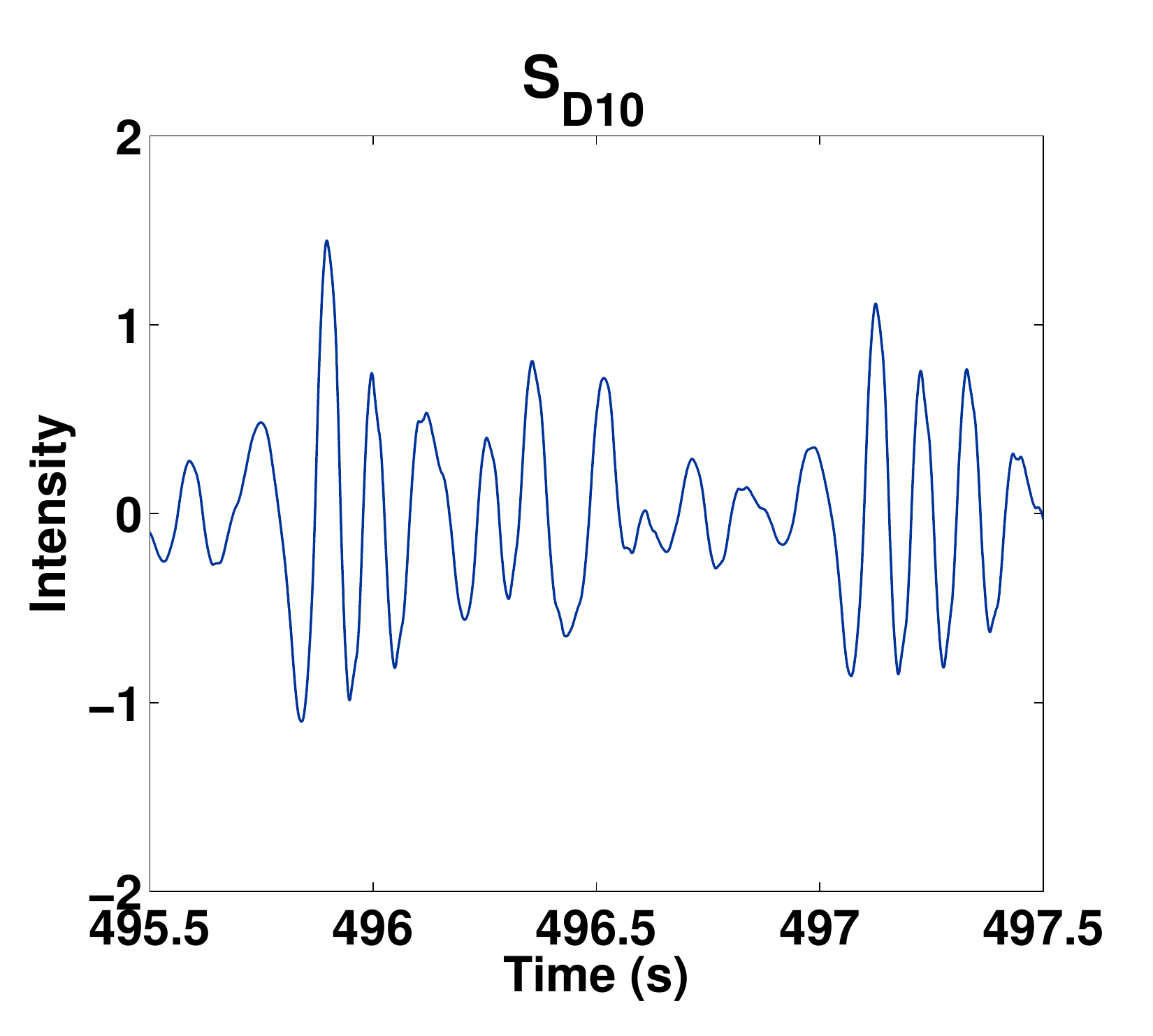}
 \end{minipage}
 \begin{minipage}[b]{0.32\textwidth}%
\centering \includegraphics[width=5.2cm,height=3.7cm]{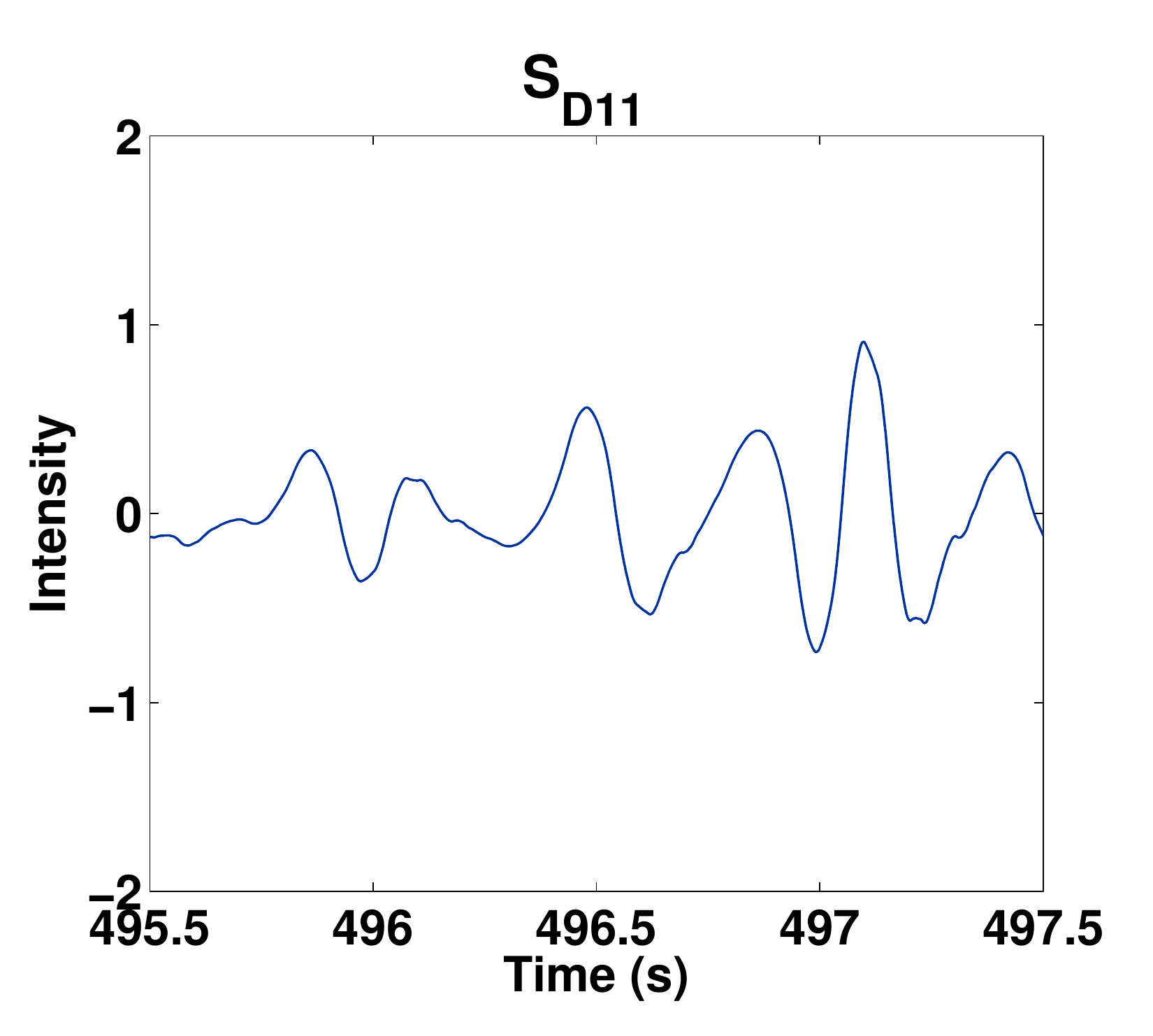}
 \end{minipage}
 \vfill
 \begin{minipage}[b]{0.32\textwidth}%
\centering \includegraphics[width=5.2cm,height=3.7cm]{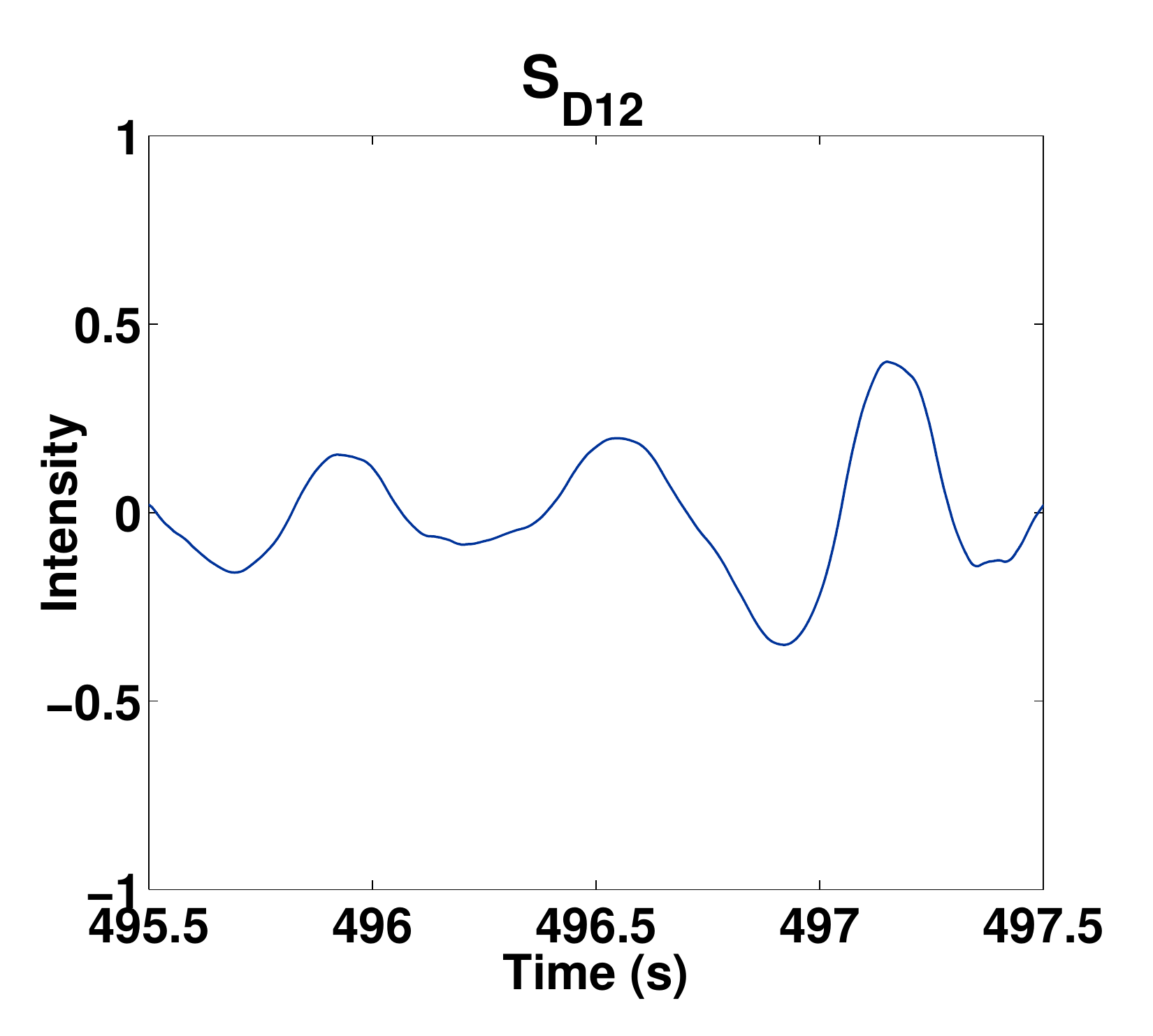}
 \end{minipage}%
 \begin{minipage}[b]{0.32\textwidth}%
\centering \includegraphics[width=5.2cm,height=3.7cm]{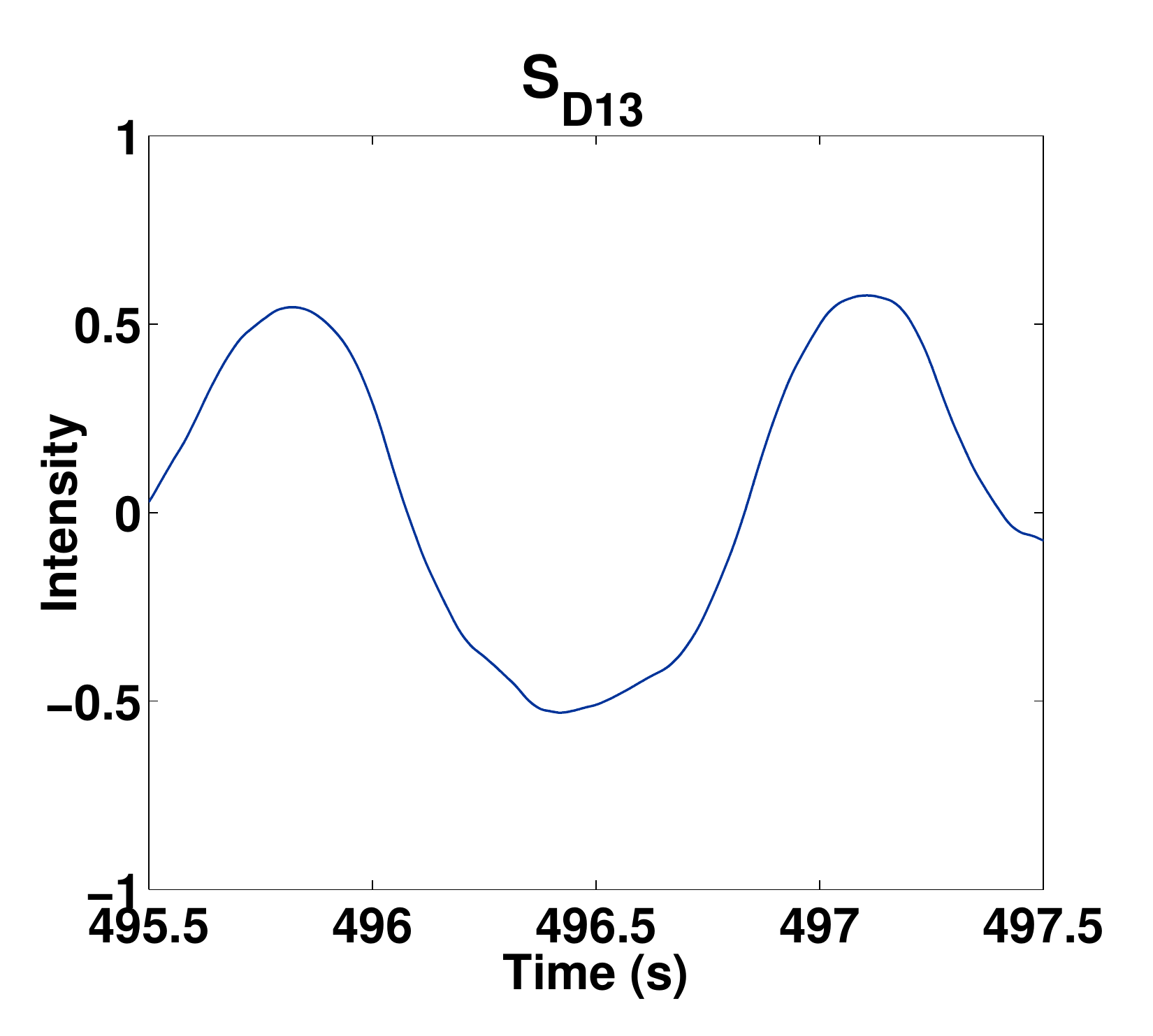}
 \end{minipage}
 \begin{minipage}[b]{0.32\textwidth}%
\centering \includegraphics[width=5.2cm,height=3.7cm]{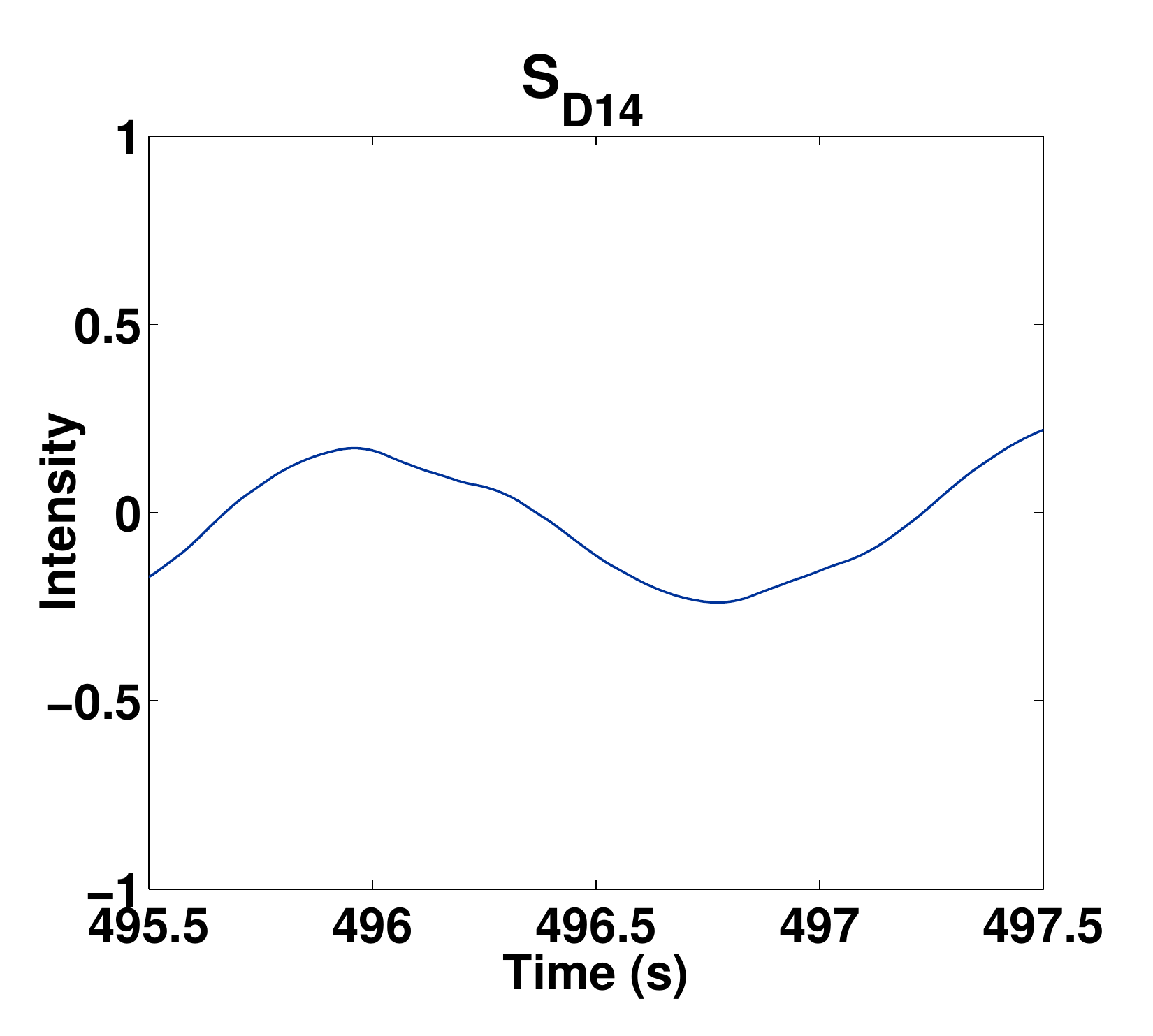}
 \end{minipage}
\vfill 
	\caption{A short slice of J1048$-$5838 de-dispersed time series in which two pulses are detected and signals reconstructed from the detail coefficients of levels $1$ through $14$. From equation (\ref{eq:cal_wavelet_select_level}), $N{_{max}}= 14$ and $ N{_{min}}= 7 $, which means levels $7$ through $14$ are expected to contribute to the signal description. Note that there are obvious pulses observed in the panels $S_{D_{7}}$, $S_{D_{8}}$ and $S_{D_{10}}$. They are the three main components that contributed to the reconstructed pulse signal. The negative intensities in the pulse signal are due to the profile of the selected mother wavelet shown in Figure \ref{fig:db5}. This artifact will be removed due to a post-processing described in Section \ref{subsec:Pulse profiles}. }
	\label{fig:wavelet_decompose}
\end{figure}

As an example, the sampling time for the observation data of J1048$-$5838 is $100$ $\mu s$ and its rotational period is 1.23 s. Therefore, $f{_{min}}=0.82$~Hz and $f{_{max}}=164$~Hz, and according to equation (\ref{eq:cal_wavelet_select_level}), $N{_{max}}=14$ and $N{_{min}}=7 $. These levels between $7$ and $14$ are expected to contribute to the description of the RRAT's pulses. Figure \ref{fig:wavelet_decompose} shows a short slice of J1048$-$5838 de-dispersed time series in which two pulses are detected and signals reconstructed from the detail coefficients of levels $1$ through $14$. There are obvious pulses observed in the panels labeled as $S_{D_{7}}$, $S_{D_{8}}$ and $S_{D_{10}}$. They are the three main components that contributed to the reconstructed pulse signal. The negative intensities in the pulse signal are due to the selected profile of the mother wavelet as shown in equation (\ref{eq:wavelet_condition}); we will show how to remove the negative components in Section \ref{subsec:Pulse profiles}. 

\subsection{Wavelet Shrinkage}  
\label{subsec:Wavelet Shrinkage}

Once the levels contributing to signal reconstruction are identified, the next step is to estimate the weights of each contributing level. We apply wavelet shrinkage, developed by \citet{dj94}, which selects wavelet coefficients based on thresholding and uses them for reconstruction. There are two types of thresholding used in practice: hard thresholding and soft thresholding \citep{bg96,fk03}. The output DWT coefficients ${W}'_{hard}(k)$ and ${W}'_{soft}(k)$ based on these two types of thresholding with threshold $\eta$ are:
\begin{equation}\label{eq:hard thresholding}
\begin{aligned}
{W}'_{hard}(k)= &
\begin{cases}
W(k), &\left | W(k) \right |> \eta \\ 
0, & \left | W(k) \right |\leqslant \eta 
\end{cases} \\
\end{aligned} 
\end{equation} 
\begin{equation}\label{eq:soft thresholding}
\begin{aligned}
{W}'_{soft}(k)= &
\begin{cases}
sign(W(k))(W(k)-\eta), &\left | W(k) \right |> \eta \\ 
0, & \left | W(k) \right |\leqslant \eta 
\end{cases}
\end{aligned} 
\end{equation} 
where $W(k)$ is the DWT coefficient before the application of thresholding. Function $sign(.)$ is defined as:
\begin{equation}
sign(x) = \begin{cases} -1, & \mbox{if } x<0 \\ 0, & \mbox{if } x=0 \\1, & \mbox{if } x>0 \end{cases}
\end{equation}

\begin{figure}[!htb]
\centering
\begin{minipage}[b]{0.32\textwidth}%
\centering \includegraphics[width=5.2cm,height=3.7cm]{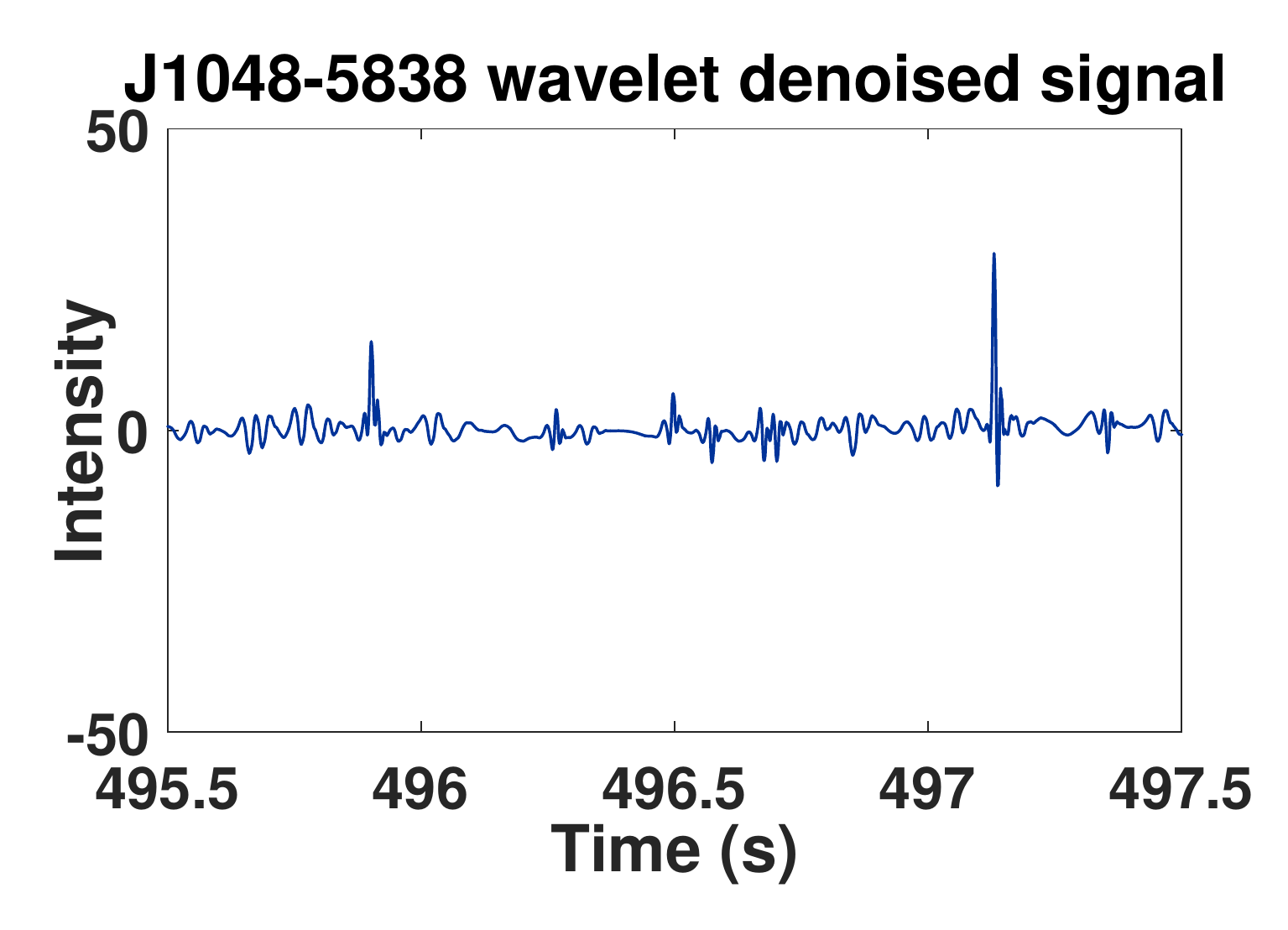}
\end{minipage}%
 \begin{minipage}[b]{0.32\textwidth}%
\centering \includegraphics[width=5.2cm,height=3.7cm]{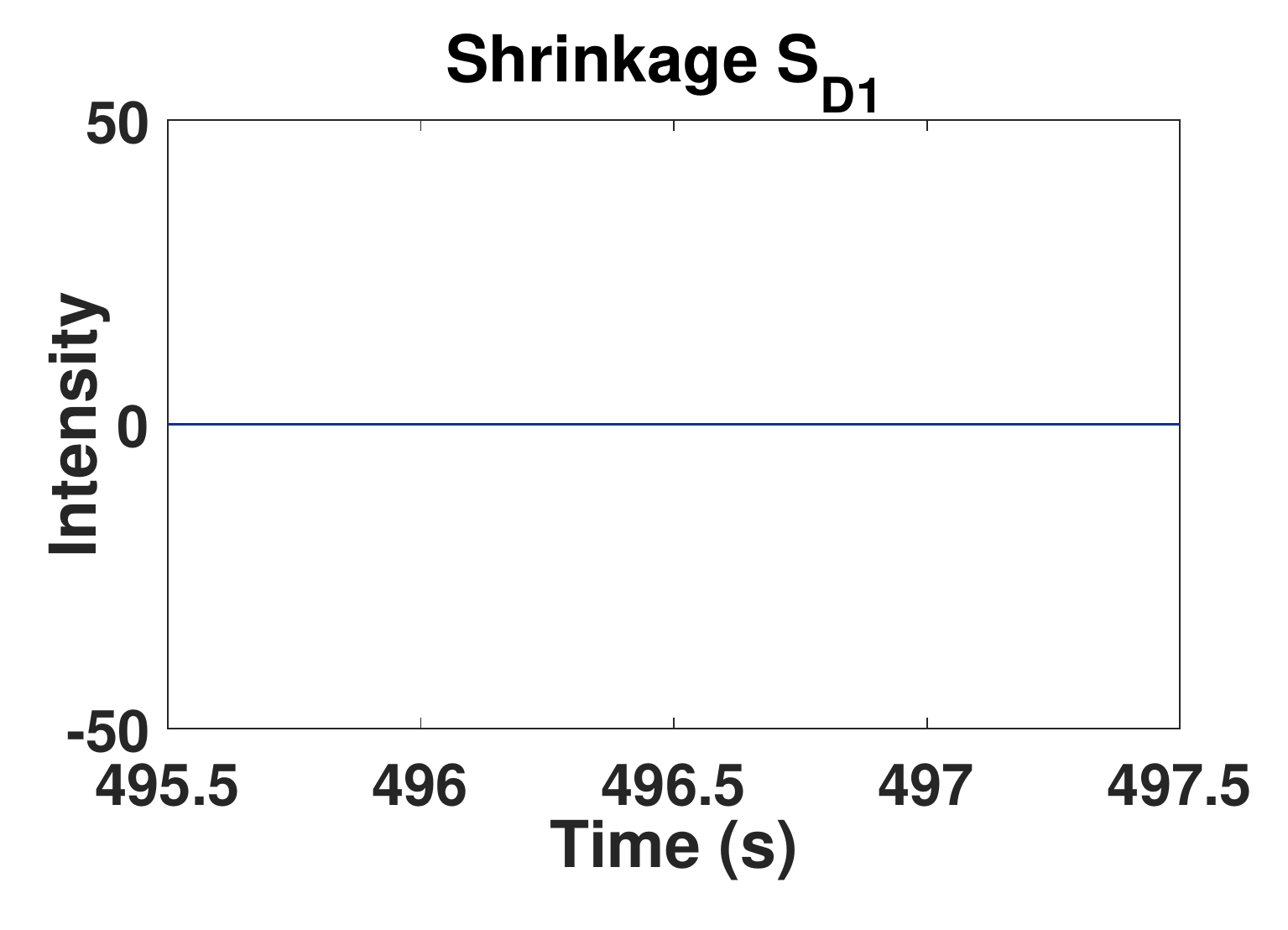}
 \end{minipage}
 \begin{minipage}[b]{0.32\textwidth}%
\centering \includegraphics[width=5.2cm,height=3.7cm]{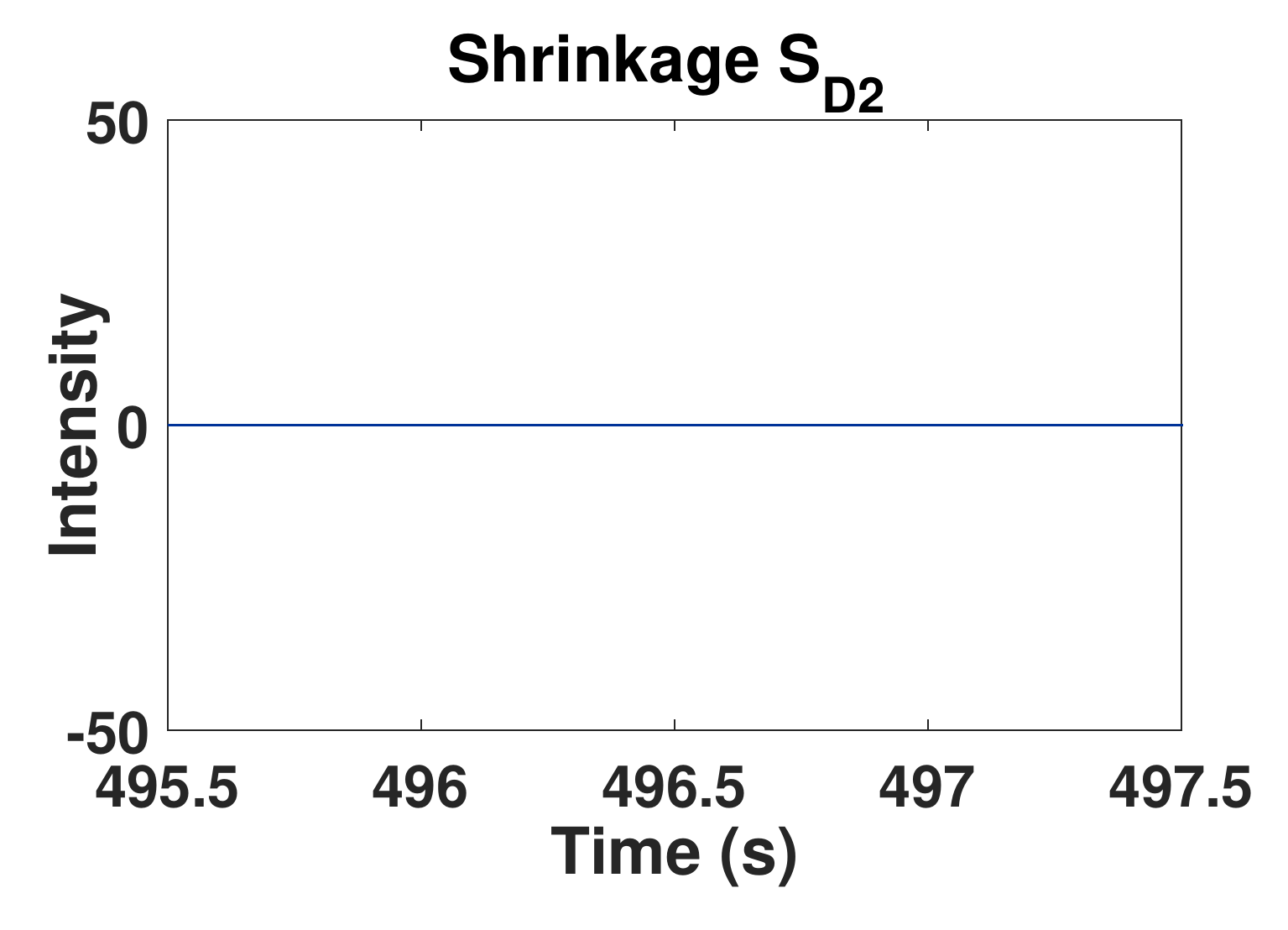}
 \end{minipage}
\vfill
\vspace{-0.15cm}
 \begin{minipage}[b]{0.32\textwidth}%
\centering \includegraphics[width=5.2cm,height=3.7cm]{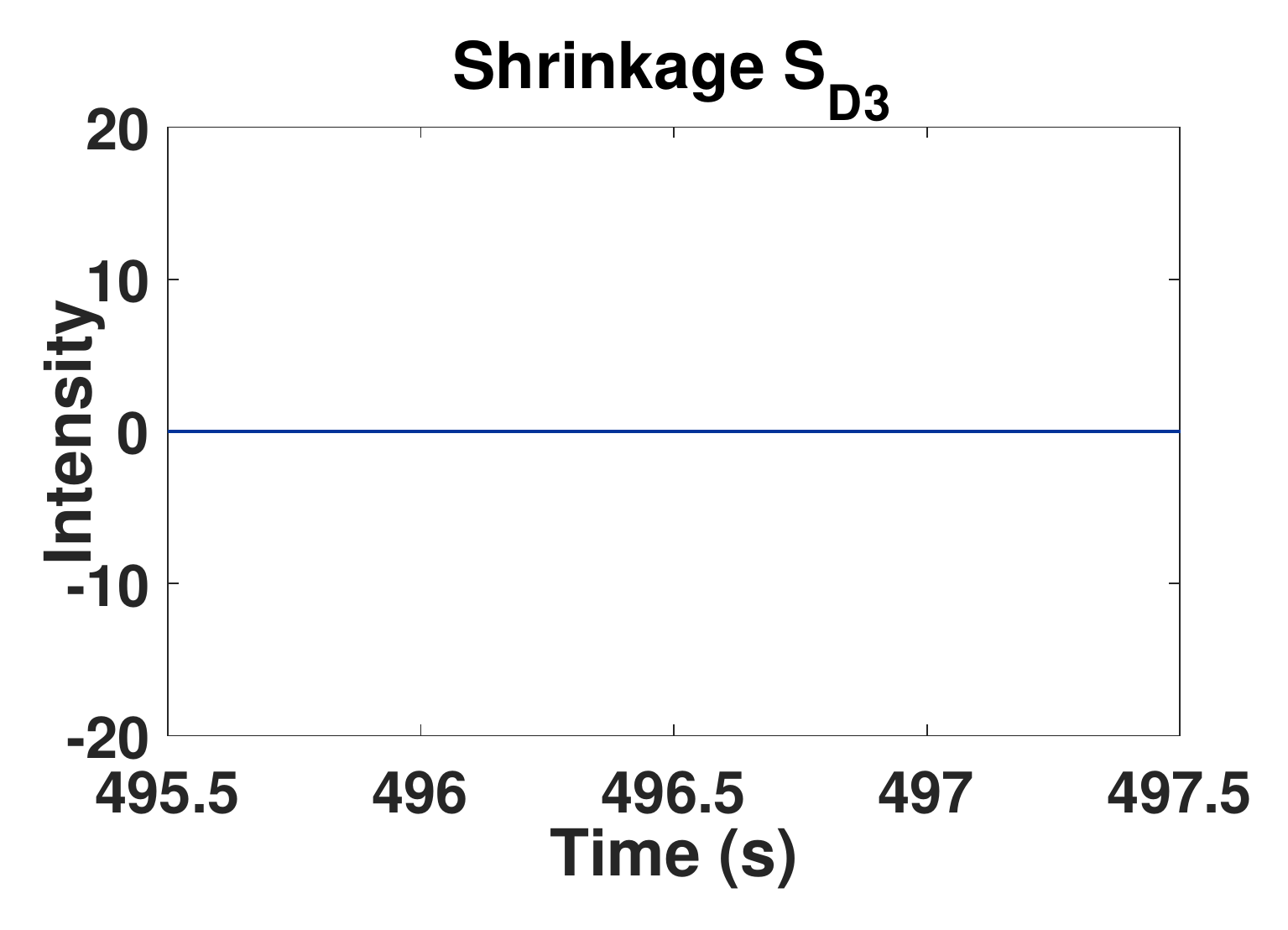}
 \end{minipage}%
 \begin{minipage}[b]{0.32\textwidth}%
\centering \includegraphics[width=5.2cm,height=3.7cm]{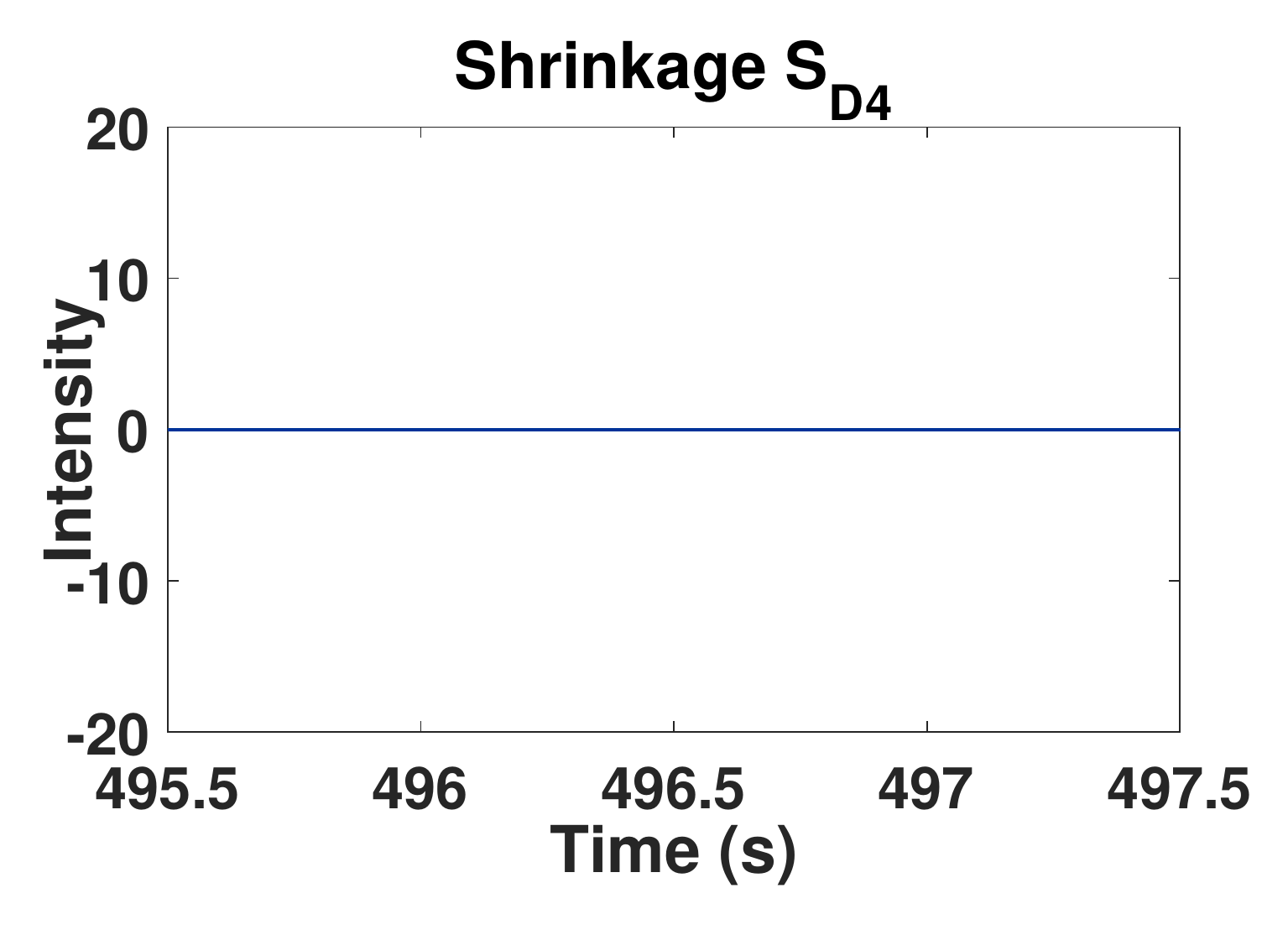}
 \end{minipage}
 \begin{minipage}[b]{0.32\textwidth}%
\centering \includegraphics[width=5.2cm,height=3.7cm]{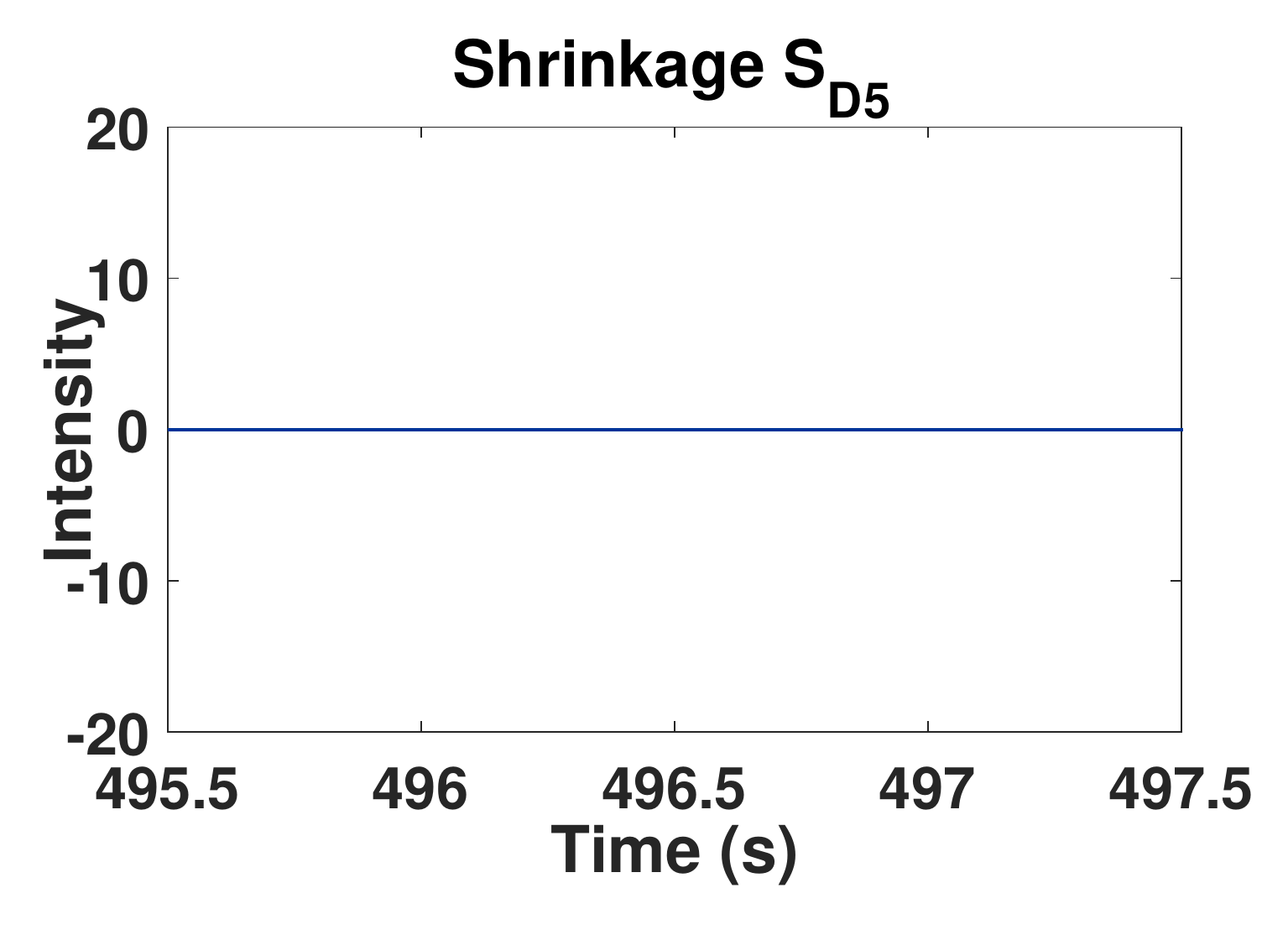}
 \end{minipage}
 \vfill
\vspace{-0.15cm}
 \begin{minipage}[b]{0.32\textwidth}%
\centering \includegraphics[width=5.2cm,height=3.7cm]{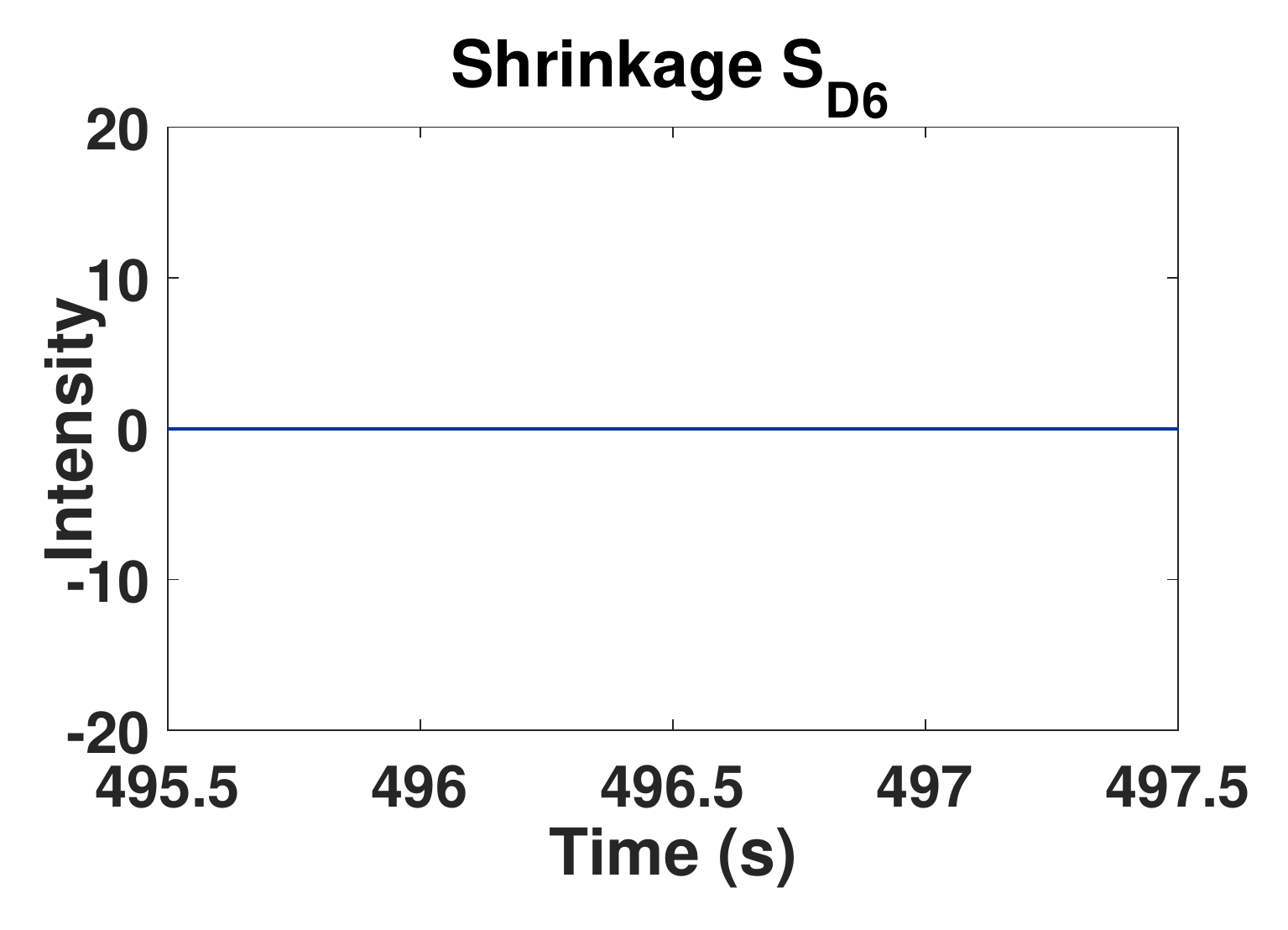}
 \end{minipage}%
 \begin{minipage}[b]{0.32\textwidth}%
\centering \includegraphics[width=5.2cm,height=3.7cm]{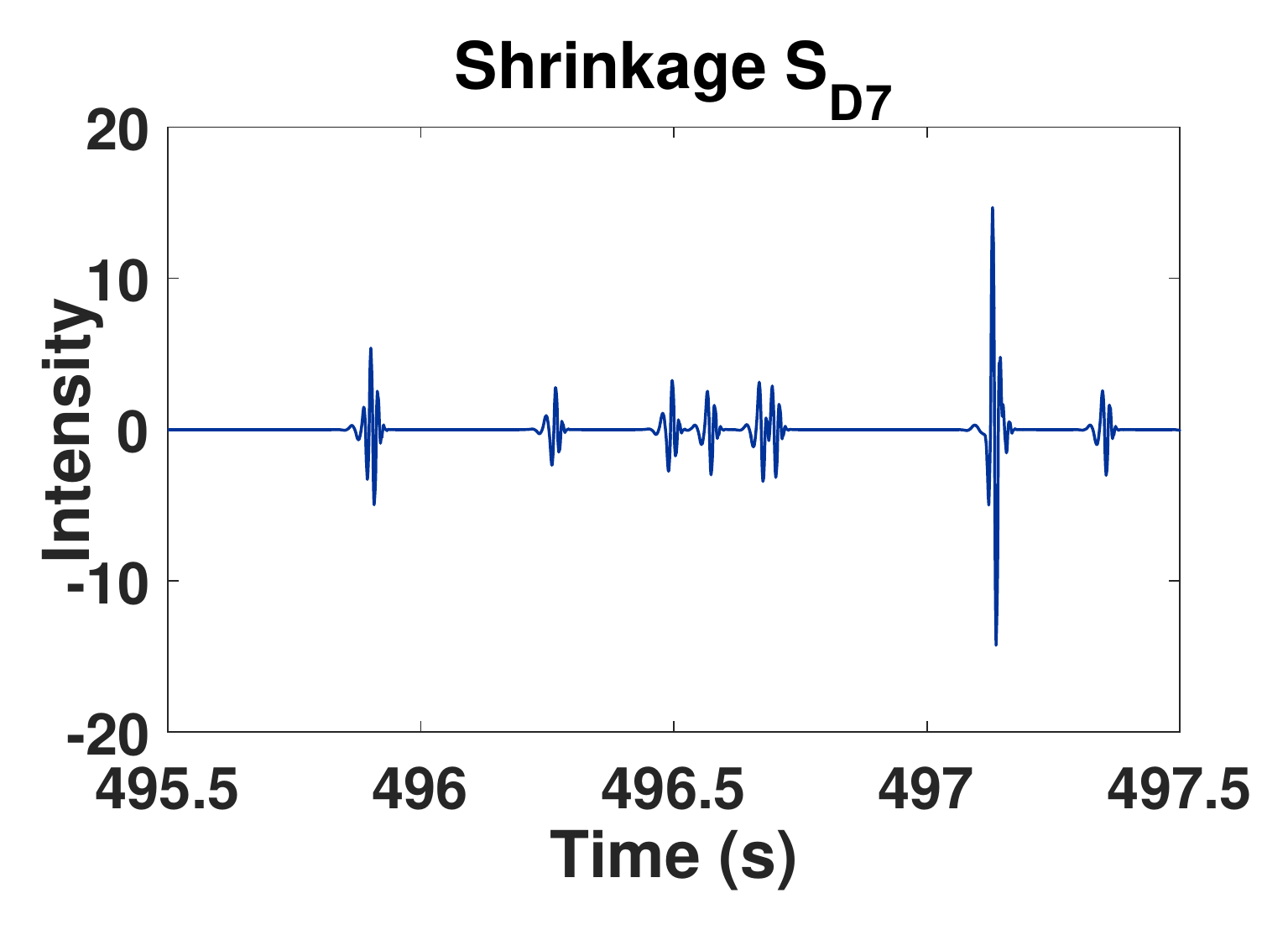}
 \end{minipage}
 \begin{minipage}[b]{0.32\textwidth}%
\centering \includegraphics[width=5.2cm,height=3.7cm]{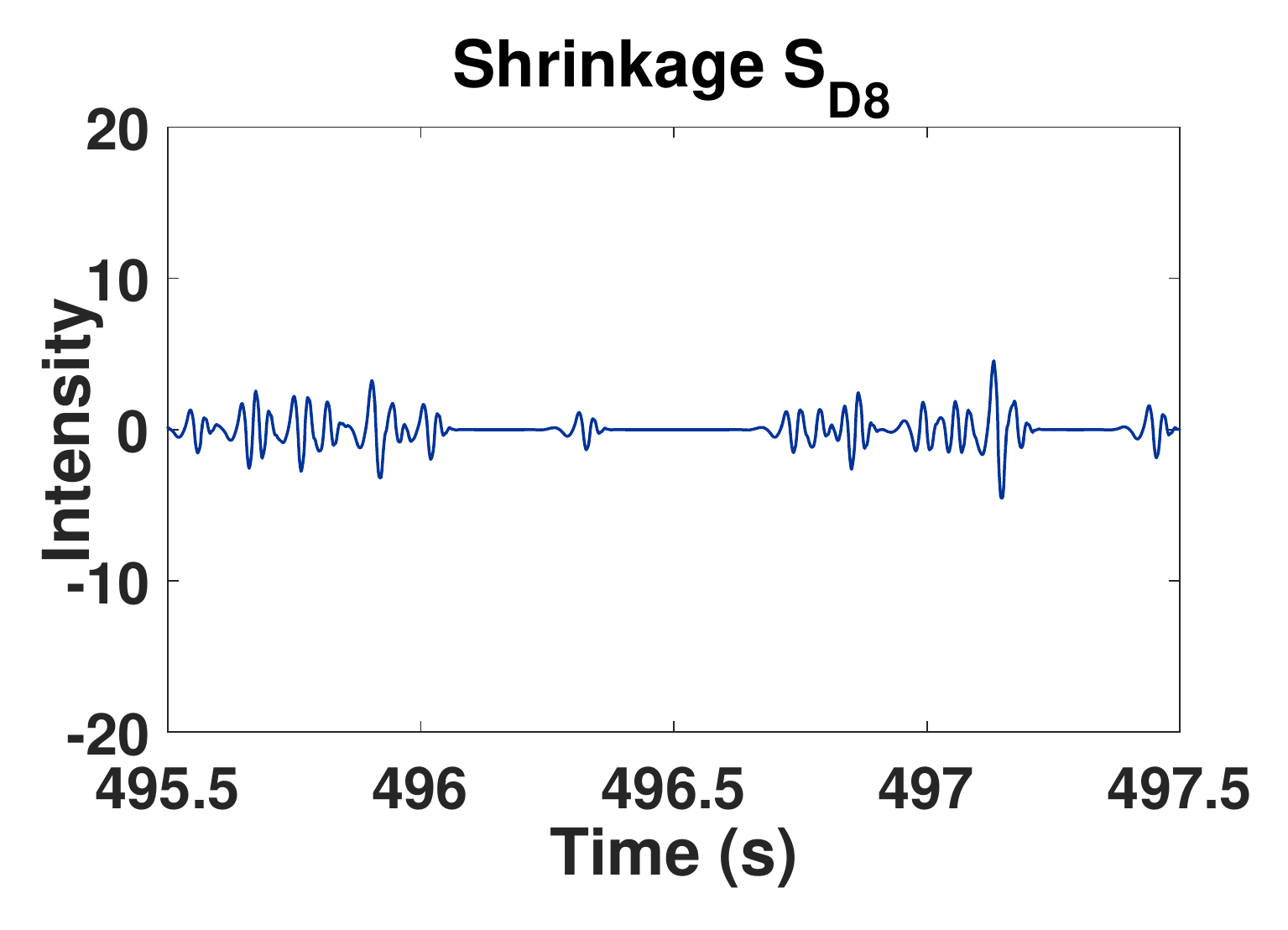}
 \end{minipage}
\vfill
\vspace{-0.15cm}
 \begin{minipage}[b]{0.32\textwidth}%
\centering \includegraphics[width=5.2cm,height=3.7cm]{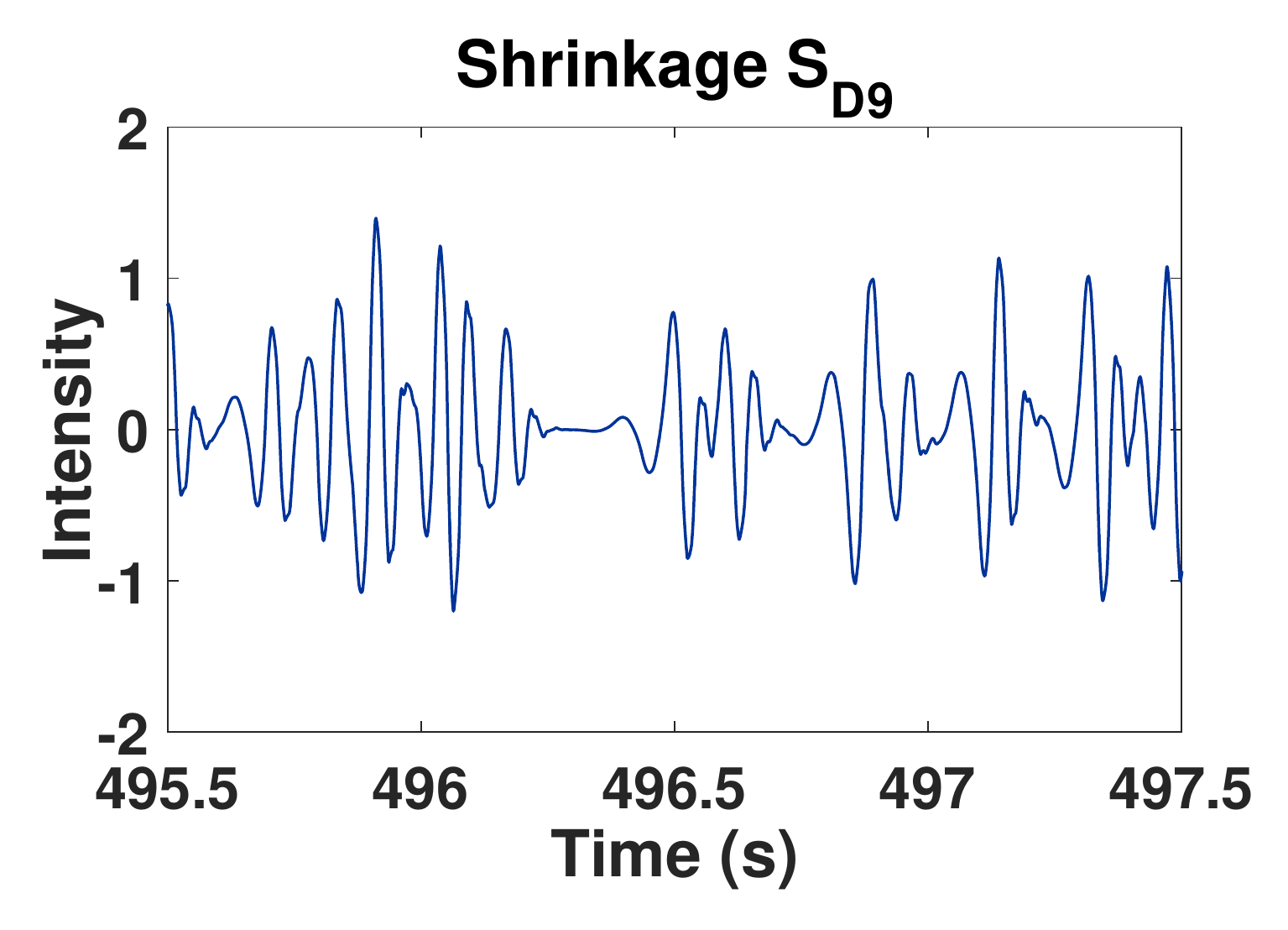}
 \end{minipage}%
 \begin{minipage}[b]{0.32\textwidth}%
\centering \includegraphics[width=5.2cm,height=3.7cm]{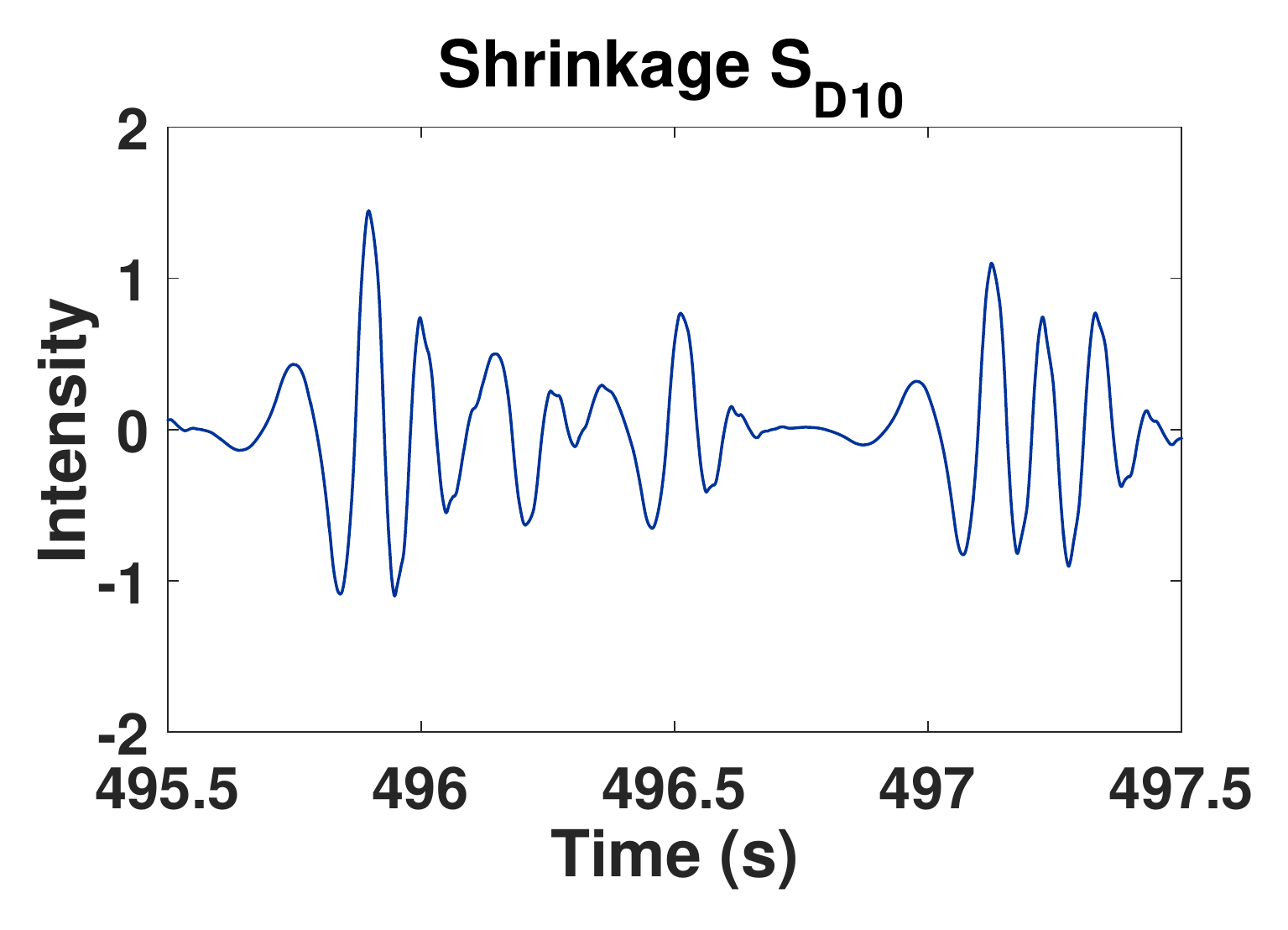}
 \end{minipage}
 \begin{minipage}[b]{0.32\textwidth}%
\centering \includegraphics[width=5.2cm,height=3.7cm]{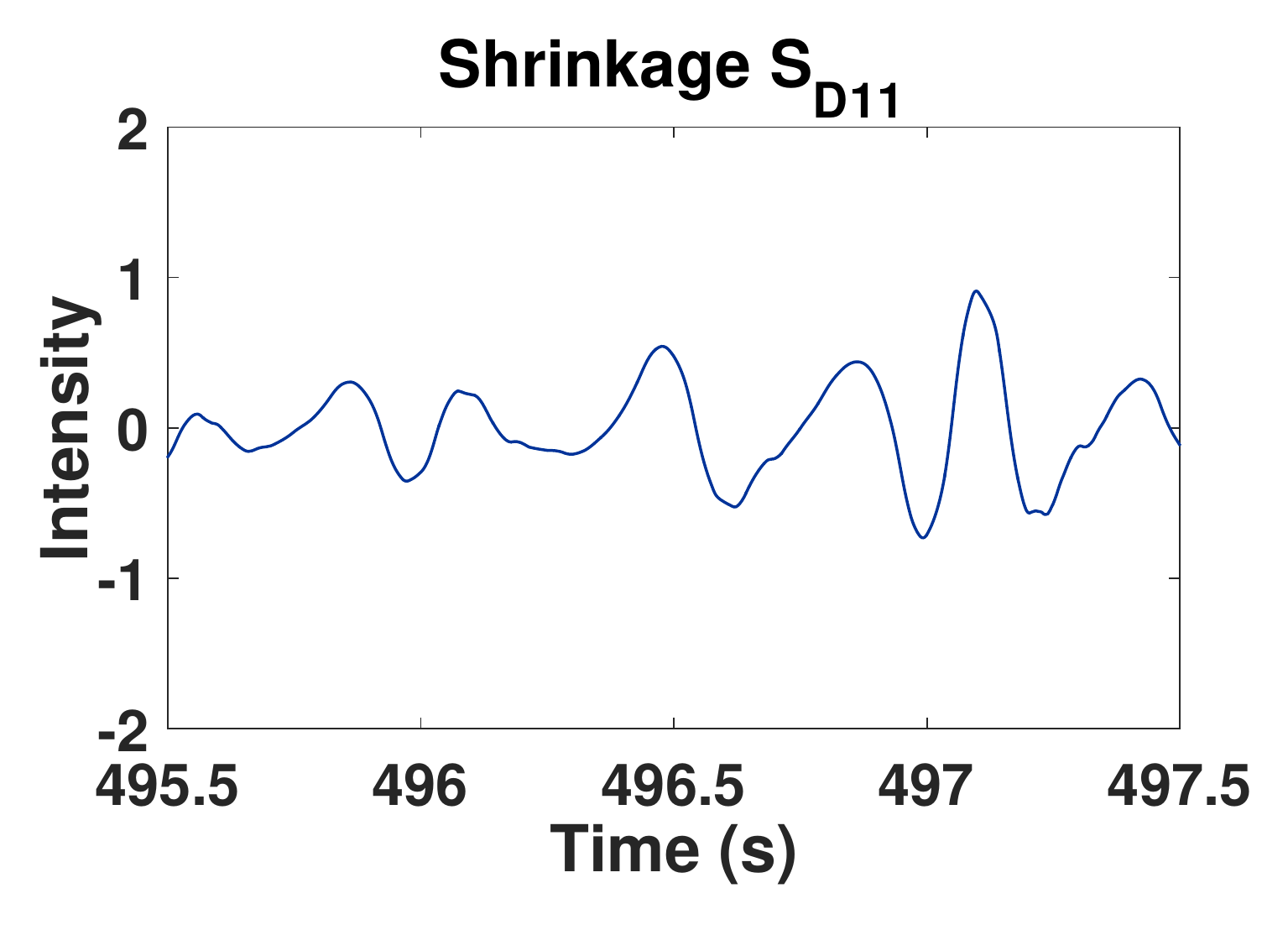}
 \end{minipage}
 \vfill
 \begin{minipage}[b]{0.32\textwidth}%
\centering \includegraphics[width=5.2cm,height=3.7cm]{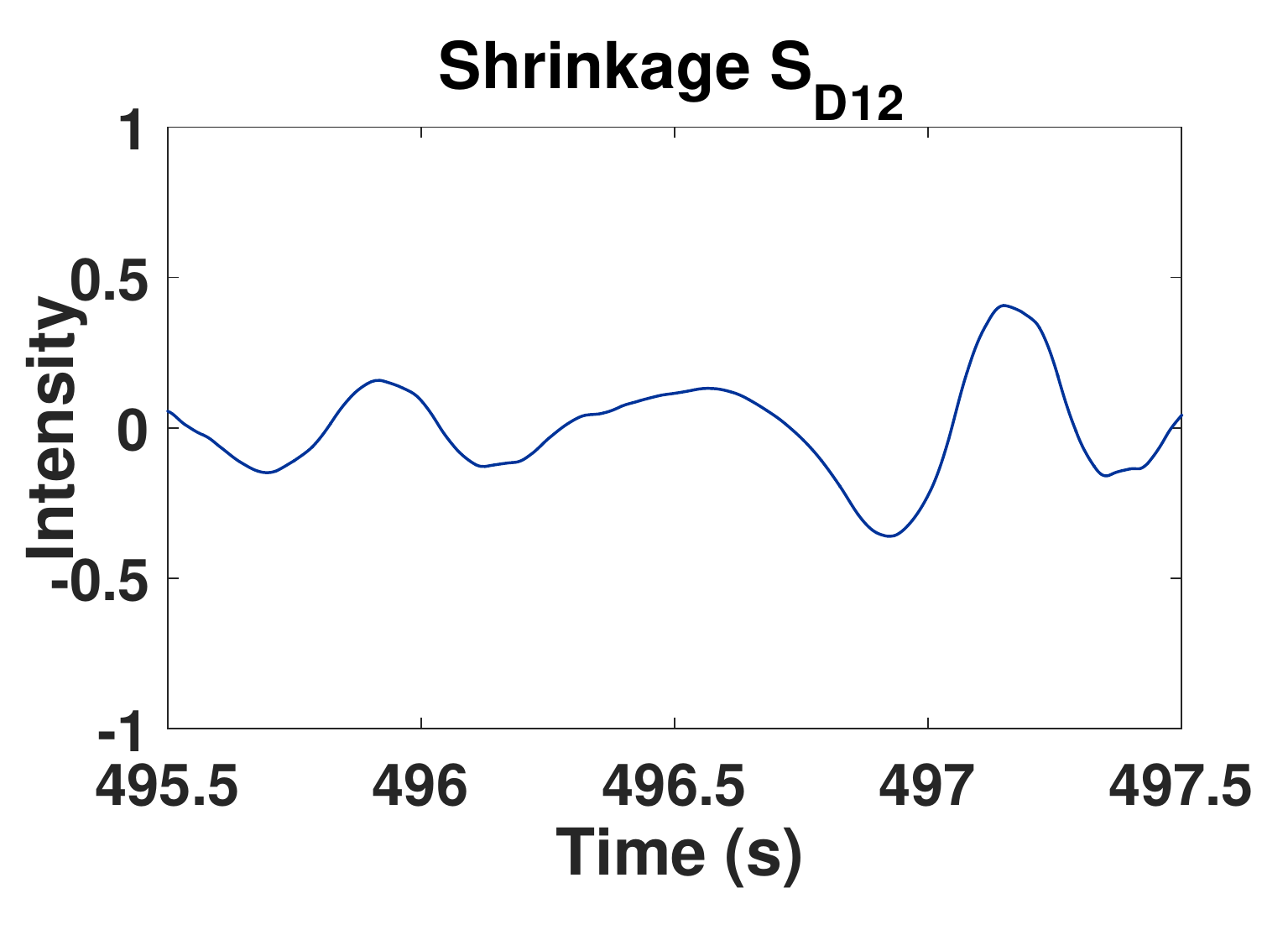}
 \end{minipage}%
 \begin{minipage}[b]{0.32\textwidth}%
\centering \includegraphics[width=5.2cm,height=3.7cm]{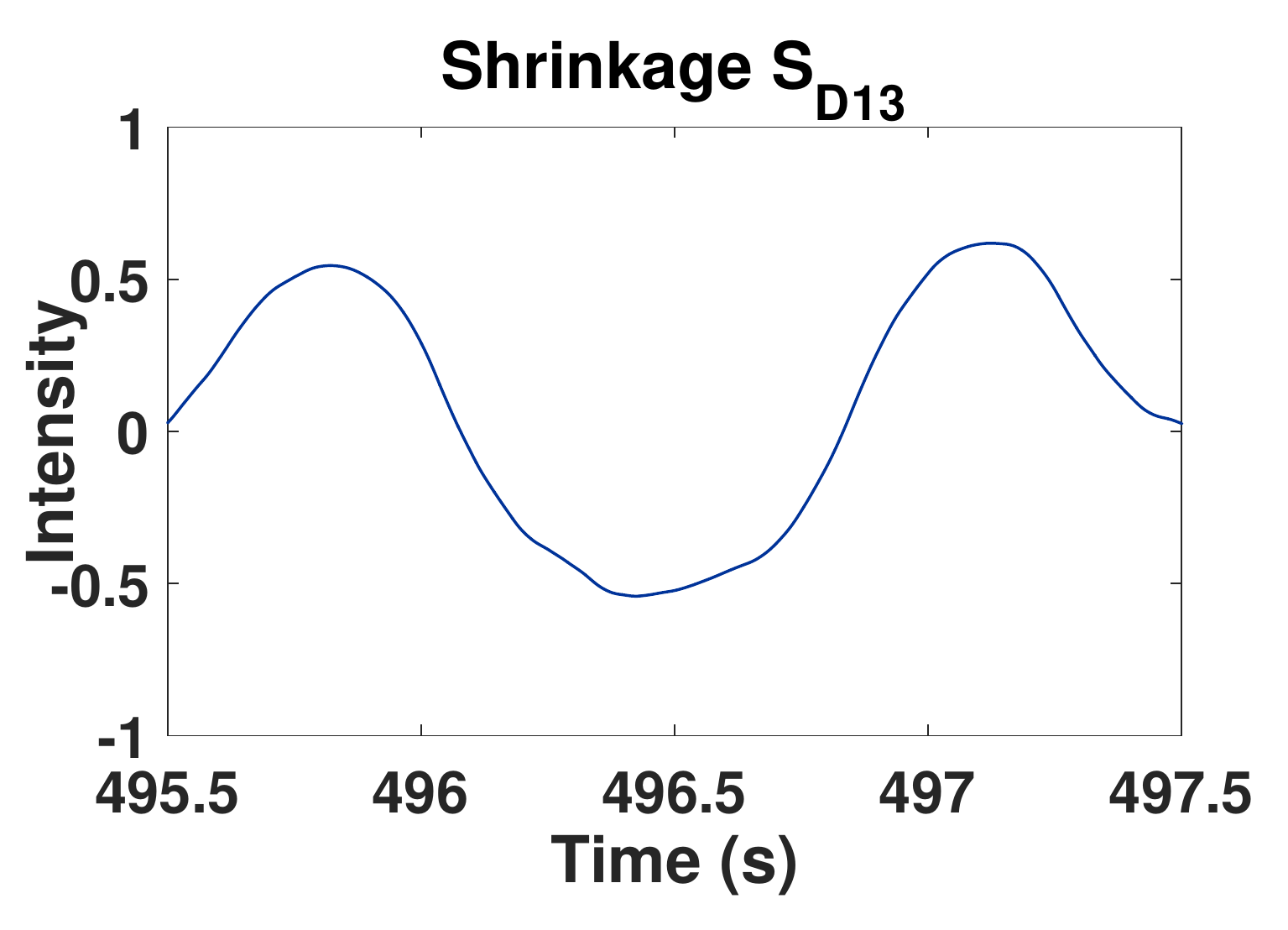}
 \end{minipage}
 \begin{minipage}[b]{0.32\textwidth}%
\centering \includegraphics[width=5.2cm,height=3.7cm]{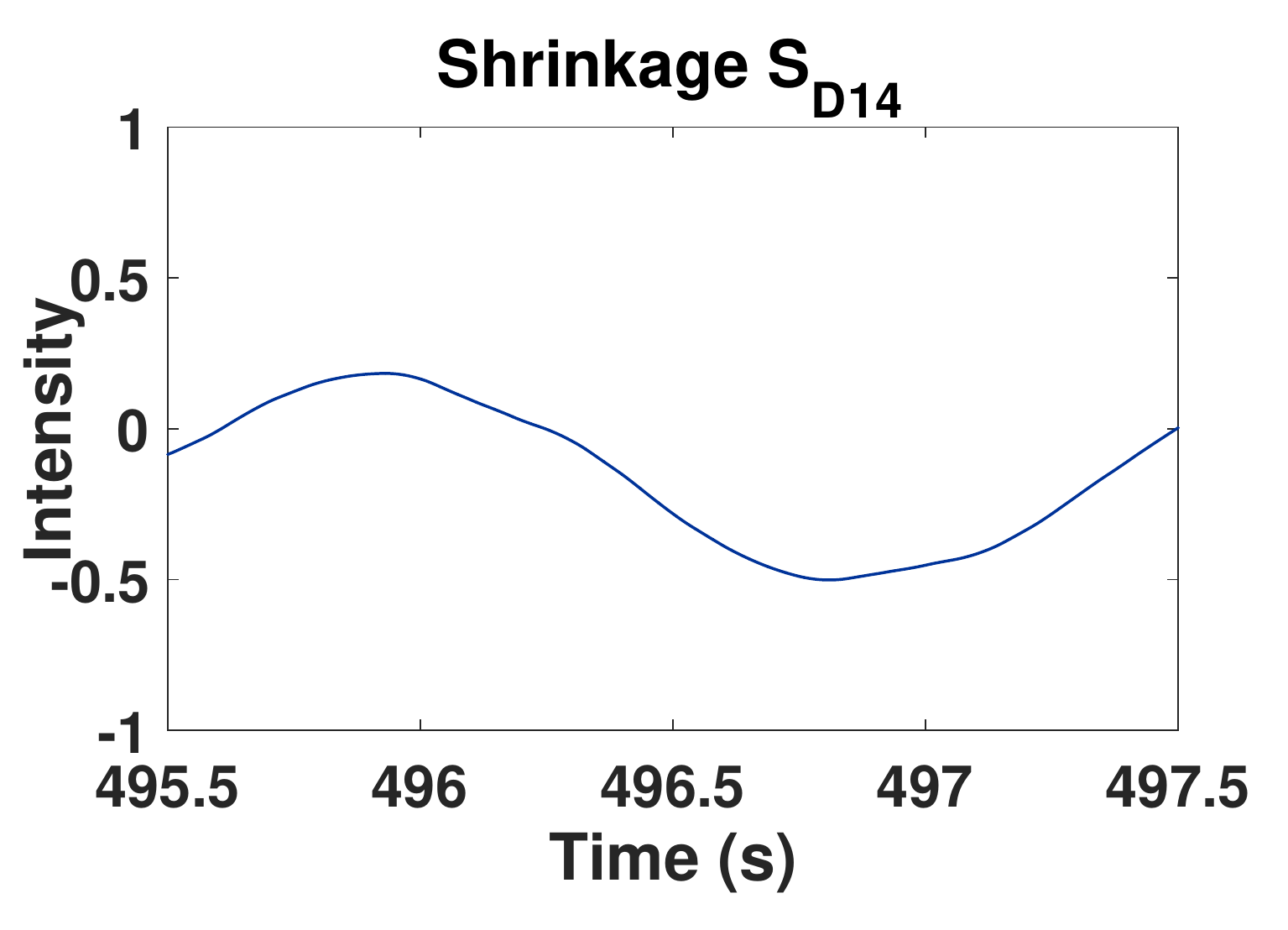}
 \end{minipage}
\vfill 
	\caption{The outcome of applying wavelet shrinkage to sub-bands $7$ - $14$ (corresponding to the wavelet decomposed levels shown in Figure \ref{fig:wavelet_decompose}.). As levels $1$ to level $6$ have no contribution to the RRAT signal, they are not used for wavelet reconstruction and are shown as a zero signal. The top left panel is the reconstructed signal after wavelet denoising. }
	\label{fig:wavelet_decompose_denoised}
\end{figure}

The observed noisy time series $Y(t)$ at time $t$ containing an RRAT signal can be modeled as an RRAT  signal $X(t)$ contaminated by an independent additive noise $N(t)$:
\begin{equation}
Y(t) = X(t) + N(t), 
\label{eq:signal_construction} 
\end{equation} 
where for a fixed value of $t$, $N(t)$ is a Gaussian random variable with mean zero and variance $\sigma^2.$ Noise samples are independent and identically distributed.  
In this paper, rule (\ref{eq:hard thresholding}) is applied to wavelet coefficients. The threshold $\eta$ is estimated based on the noise variance in each sub-band. In Figure \ref{fig:wavelet_decompose}, the signal is decomposed into 14 levels (sub-bands), where levels $7$ through $14$ are expected to contribute to the RRAT signal description. Figure \ref{fig:wavelet_decompose_denoised} shows the outcome of applying wavelet shrinkage to sub-bands $7$ - $14$. As levels $1$ to level $6$ have no contribution to the RRAT signal, they are not used for wavelet reconstruction and are shown as a zero signal. The top left panel is the reconstructed signal after wavelet denoising. Comparing with the original signal in Figure \ref{fig:wavelet_decompose}, we can see the noise is suppressed.

\begin{figure}[!htb] 
\centering
\vspace{-0.3cm}
\includegraphics[width=0.5\textwidth]{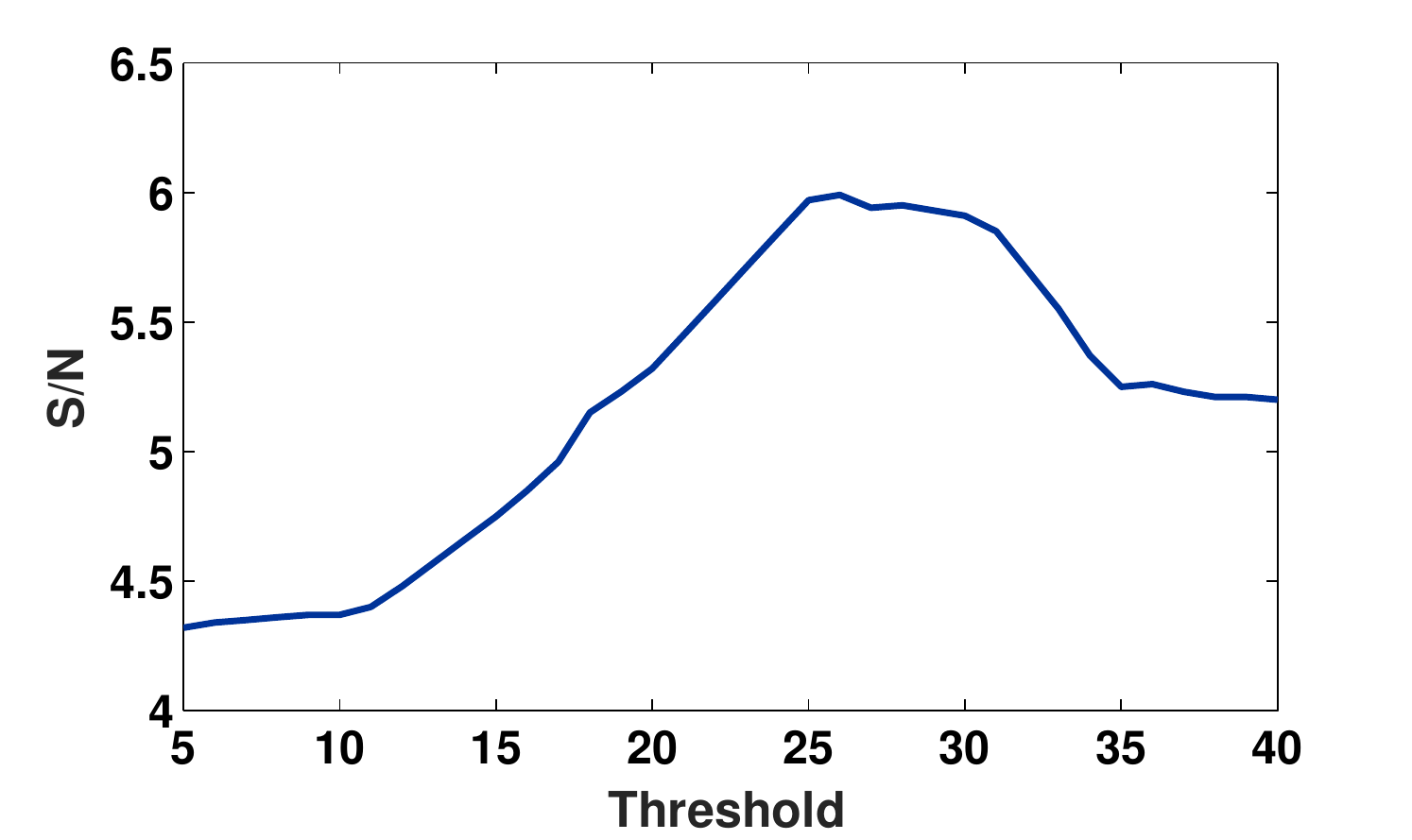} 
\vspace{-0.15cm}
	\caption{Shown is the dependence of the S/N value on the threshold $\eta$ in the process of selecting the weight of the detail coefficients at level 7 for J1048$-$5838's time series. The x-axis is the threshold $\eta$ applied to the wavelet coefficients, the y-axis is an S/N of the pulse after thresholding. Here, $\eta$ is estimated based on the noise variance in each sub-band. }
	\label{fig:threshold_SNR}
\end{figure}

Figure \ref{fig:threshold_SNR} shows the dependence of the S/N value on the threshold $\eta$ in the process of selecting the weight of the detail coefficients at level $7$ for J1048$-$5838's time series. Note that S/N peaks at $\eta = 25,$ which is selected as the optimal threshold at level $7$. RRATs have various optimal $\eta$ for different sub-bands.

In this paper, we first apply wavelet analysis (decomposition) to time series containing an RRAT signal, then select the decomposition levels over the range of frequencies containing the RRAT signal and apply optimal thresholds to the detail coefficients at the selected levels. The last step is the reconstruction of the signal from the coefficients of the selected levels. We use the Wavelet Analysis toolbox from Matlab \citep{Matlab} for wavelet decomposition and thresholding. We use the \textit{wavedec()} function for multilevel wavelet decomposition, the function \textit{wthcoef()} for the wavelet coefficients thresholding, and the \textit{wrcoef()} function for reconstructing single pulses from the wavelet coefficients. 
  
\section{Performance Measures} 
\label{sec:Analysis}

We compare the conventional denoising method currently used in single-pulse searching$-$boxcar smoothing \citep{cm03} with the wavelet denoising approach. After applying wavelet denoising to a time series, we detect candidate pulses with peak S/N exceeding a preset threshold. The peak S/N is calculated from the single-pulse profile and is defined as the peak divided by the root mean square of the noise (the profile amplitude in the off-pulse region). The S/N threshold is typically around 5$\sigma$ or 6$\sigma$, where $\sigma$ is the root mean square of the noise in each period. In this paper, the S/N threshold is set to either 4.5$\sigma$ or to  5$\sigma$, depending on the data. This indicates that we are able to detect pulses from very faint sources. After identifying candidate pulses, we proceed with the periodicity check. We use temporal locations of peaks as the times of arrival. If the times of arrival $t_1$ and $t_2$ of an arbitrary pair of pulses satisfy the following inequality 
\begin{equation}\label{eq:periodicity check} 
\left | \frac{t_{1}-t_{2}}{P}-round\left(\frac{t_{1}-t_{2}}{P}\right) \right| <\varepsilon,  
\end{equation}
then the pulses are retained for further analysis. Here, $P$ is the spin period of an RRAT and $\varepsilon$ is a tolerance for the periodicity check. 

We use False Alarm Rate (FAR) as a measure of detection performance. Here FAR is defined as the fraction of pulses assumed to be real, which may actually occur due to RFI accidentally falling into the periodicity window ($2\varepsilon$). It is defined as: 
\begin{equation}
{\rm{FAR}}=\frac{(N_{candidates}-N)\cdot 2\varepsilon }{N},
\label{eq:far} 
\end{equation}
where $N_{candidates}$ is the number of pulses after denoising and $N$ is the number of pulses remaining after the periodicity check. 

At last, to determine which pulses are due to an RRAT and which are spurious, we perform a timing analysis, as described in \citet{cbm+17}. We delete data points with large residuals outside of a reasonable range of jitter (generally the pulses with residuals that are larger than the average TOA error), which produces errors due to the intrinsic pulsar emitting mechanism, or intrinsic inconsistencies in pulse phase. The number of real pulses determined by this method is listed in the sixth column in Table \ref{tab2} .

\section{Numerical Results} 
\label{sec:Results}

In this section, we first present the result of wavelet denoising applied to the time series of eight RRATs. Then we compare two timing solutions - one using TOAs calculated from data that have been denoised using the wavelet approach and the other using the conventional method (boxcar) in \citet{cbm+17}. Along with the comparison of the timing solution, we compare the shape of denoised RRAT pulses. 

\subsection{Denoising} 
\label{subsec:Denoising results}

Smoothing data with a boxcar filter is a conventional denoising approach used for pulsar radio observation data. It lowers the effective sampling rate of the time series by binning the data (resampling the data with a lower sample rate), typically using 128, 256, or 512 bins per period. The boxcar filter is usually applied in two stages. In the single-pulse search stage (without knowing the pulsar spin period), an $N$-sample boxcar (the width of boxcar is $N$ times the sampling interval, here $N$ is optimized by maximizing the S/N) is applied to the original time series to identify the pulse candidates. Then another boxcar (the width of boxcar is different from the one utilized at the previous stage, generally 128, 256, or 512 bins per period) is applied to time series for calculating the profile of the pulse. 

\begin{figure}[!t]
\centering
\vspace*{-0.05cm}
\includegraphics[width=8cm]{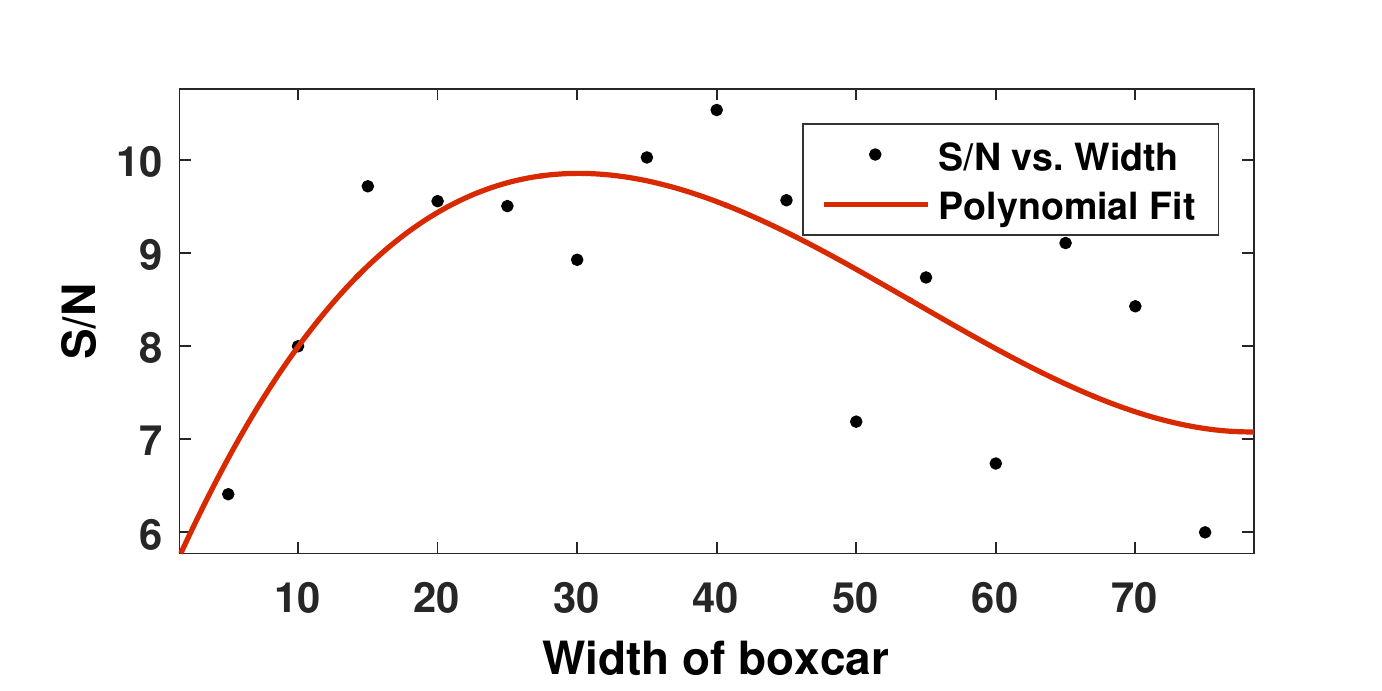}\\
\vspace*{0.17cm}
\includegraphics[width=8cm]{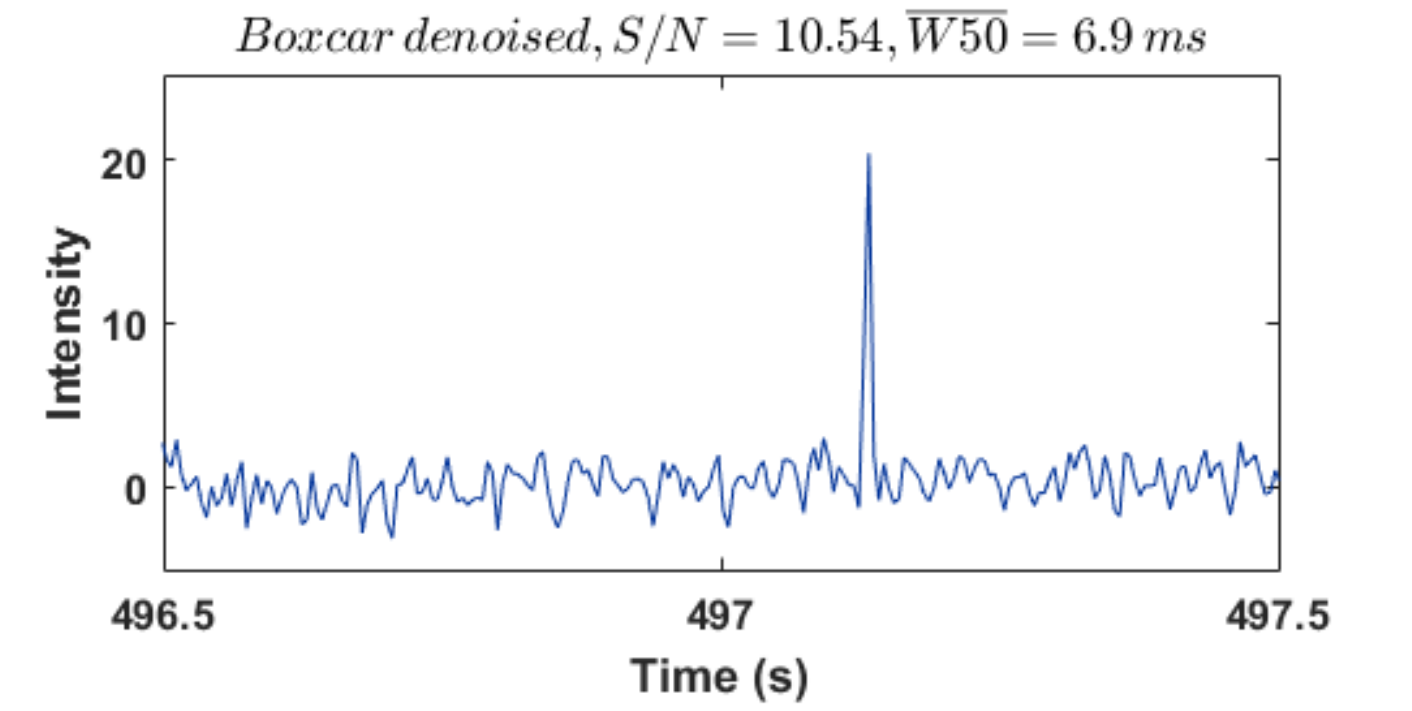}\\
\vspace*{0.16cm}
\includegraphics[width=8cm]{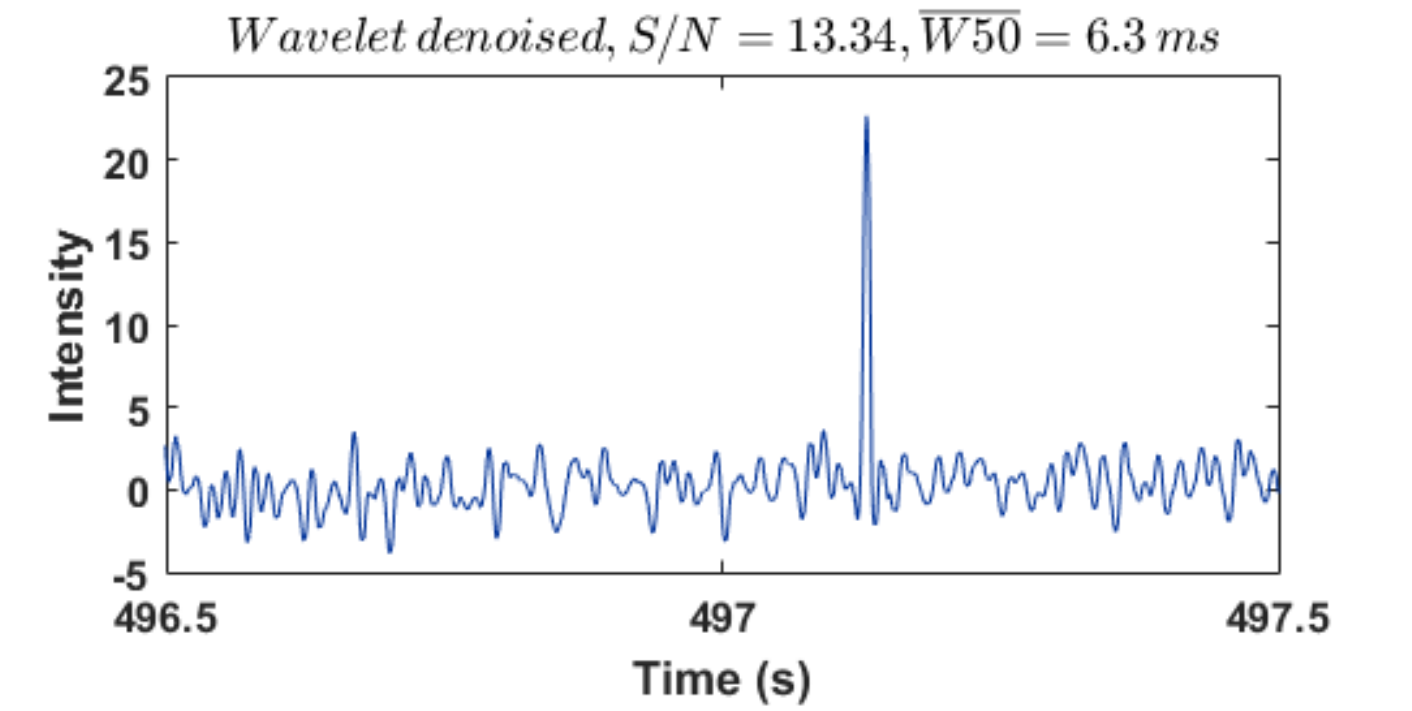}\\
\vspace*{-0.05cm}
	\caption{The comparison between a boxcar denoised profile and the same pulse denoised with wavelets on a short slice of a J1048$-$5838 time series that contain a strong pulse. The top panel shows how we select the optimal width of the boxcar for these data. The x-axis is the width of the boxcar used for denoising, and the y-axis is the corresponding S/N of the pulse. Here, we use 30 time samples as the optimal width of the boxcar. The middle panel shows the boxcar denoised data and the panel at the bottom shows the wavelet denoised data. Here, the S/N of the pulse is improved from 10.54 to 13.34. } 
	\label{fig:wavelet_compare}
\end{figure}

On the contrary, wavelet denoising will not decrease the sampling rate of the time series. The comparison between the results of denoising with wavelets and boxcar applied to time series of eight RRATs is given in Table \ref{tab2}. Figure \ref{fig:wavelet_compare} shows a comparison between boxcar denoised data and wavelet denoised data on a short slice of a J1048$-$5838 time series, which contains a strong pulse. The first panel shows how we obtain the optimal width of the boxcar for these data. The x-axis is the width of the boxcar used for denoising, and the y-axis is the S/N of the pulse for the corresponding width of boxcar. The red curve is the polynomial fitting curve. Here, we use 30 time samples as the optimal width of the boxcar. However, in the pulse detection stage, it is hard to guarantee that the optimal boxcar is applied to every pulse, as they are applied in discrete steps. The second panel shows boxcar denoised data and the last panel shows the wavelet denoised data. The S/N of the pulse is improved from 10.54 to 13.34. Such improvement in S/N is particularly meaningful for weak pulses, as they may be undetectable in the case of boxcar denoising and become detectable after denoising with wavelets. This is why most of the newly found pulses in the eight time series have  S/N around five. We will further discuss the distribution of S/N in Section \ref{sec:Discussion}. The width of the pulse ($\overline{W50}$) is calculated at 50\% of the peak intensity. The width of the pulse is labeled in the figure, with pulses in data denoised with wavelets appearing narrower compared to the same pulses in the data denoised with the boxcar.

\begin{table*}[!t]
\vspace{0cm}
\centering
\resizebox{\textwidth}{!} {%
\hspace{-3cm}
\begin{tabular}{lccccccccc}
\tableline\tableline
PSR Name&Number of Pulses After&Number of Pulses After&Tolerance&FAR&Number of&Number of Real Pulses&Increase in&Number of&Additional  \\
&Denoising ($N_{candidates}$)&Periodicity Check ($N$)&$\left ( \varepsilon \right )$&$\left ( \% \right )$&Real Pulses&with Boxcar Denoising&Detected Pulses&Epochs for Fitting& Epochs\\
\tableline
J0735$-$6302  & 1613 & 497 & 0.01  & 2.25  &  479 & 304 & $58\%$ & 6 & 1 \\
J1048$-$5838  & 799 & 374 & 0.01 & 2.41 & 310 & 207 & $51\%$ & 24 & 5\\
J1226$-$3223  & 1419 & 469 & 0.01 & 4.05 & 430 & 360& $19\%$ & 15 & 0\\
J1623$-$0841  & 7704 & 1693 & 0.01 & 7.10 & 1513 & 1202& $26\%$ & 16 & 0\\
J1739$-$2521  & 1955 & 580 & 0.015 & 7.20 & 543 & 321& $69\%$ & 16 & 2\\
J1754$-$3014  & 2460 & 862 & 0.03 & 11.12 & 829 & 550& $51\%$ & 19 & 2\\
J1839$-$0141  & 1651 & 500 & 0.01 & 4.80 & 481 & 386& $25\%$ & 14 & 1\\
J1848$-$1243  & 2133 & 522 & 0.015 & 9.26 & 440 & 393& $12\%$ & 20 & 1\\
\tableline
\end{tabular}}
\caption{Wavelet denoising results for time series of eight RRATs based on the observations in Table \ref{tab1}. Note. The second column of Table \ref{tab2} lists the number of pulses detected in the time series after the data were processed with wavelets. The third column in the table is the number of pulses remaining after the periodicity check. The fourth column shows the tolerance for the periodicity check. The fifth column lists FARs for the eight RRATs. Here FAR is the fraction of pulses assumed to be real which may actually be due to RFI falling into the periodicity window ($2\varepsilon$) randomly. The sixth column summarizes the number of remaining pulses that are determined as real and used for pulsar model timing. The seventh column lists the number of detected pulses when the boxcar denoising method has been effectively applied. The eighth column presents the difference of the results in the sixth and seventh columns, which corresponds to the difference in the detected pulses in the case of the wavelet based and boxcar denoising. Note that the number of correctly detected pulses for the case of wavelet denoising is higher compared to the case of denoising with the conventional boxcar filter. The ninth column lists the number of epochs used for pulsar parameter fitting. The last column is the number of additional epochs which had zero detected pulses based on boxcar denoising, but have a few detected pulses with wavelet denoising. \label{tab2}}
\end{table*}

\subsection{Timing Solution} 
\label{subsec:Timing solution}

We calculate TOAs for the RRAT pulses by calculating the time of the peak of the pulse. We do this instead of cross-correlating with a template, due to the variable single-pulse shapes. We then use the pulsar timing TEMPO2 \citep{hem06} software to fit a timing model by minimizing the chi-square criterion of the timing residuals. The results are listed in Table \ref{tab3}. Comparing with the previous results based on the conventional boxcar approach given in \citet[see][Table 2]{cbm+17}, which is shown in Table \ref{tab4}, we see that most of the solutions exhibit smaller errors of right ascension (R.A.), declination (decl.), and period derivative ($\dot{P}$), but larger errors in period. We will discuss this in more detail in Section 6.2. 

\begin{table*}[!t]
\vspace{0cm}
\centering
\resizebox{\textwidth}{!}{%
\hspace{-3cm}
\begin{tabular}{lccccccccccc}
\tableline\tableline
PSR Name & R.A. & Decl. & $P$ & $\dot{P}$ & RMS & DM & Epoch & Width & $B$ & $\dot{E}$ & $\tau$\\
 & (J2000) & (J2000) & (s) & ($10^{-15}$ s s$^{-1}$) & (ms) & (pc cm$^{-3}$) & (MJD) & (ms) & ($10^{12}$ G) & ($10^{31}$ergs s$^{-1}$) & (Myr)\\
\tableline
J0735$-$6302 & 07:36:20.01(27)	& $-$63:04:16(2) 		& 4.8628739612(7)		& 151.89(25) 	& 6.32 	& 19.4 		& 56212 	& 30 	& 27.5	& 5.2 	& 0.5\\
J1048$-$5838 & 10:48:12.561(14)	& $-$58:38:19.02(10) 	& 1.231304776630(4) 	& 12.19375(7) 	& 2.47 	& 70.7(9) 		& 53510	& 7   & 3.9 	& 26 	& 1.6\\
J1226$-$3223 & 12:26:46.628(40)	& $-$32:23:01(1) 		& 6.1930040852(5) 		& 7.05(1)   	& 10.08 	& 36.7 		& 56114	& 34	& 6.7 	& 0.12 	& 13\\
J1623$-$0841 & 16:23:42.6827(97) 	& $-$08:41:36.6(5) 		& 0.50301500560(1) 		& 1.9556(7)	& 3.74 	& 59.79(2) 	& 55079	& 11	& 1.0 	& 61 	& 4.1\\
J1739$-$2521 & 17:39:32.629(54)	& $-$25:21:56(15) 		& 1.81846119286(24) 	& 0.24(2)		& 19.17 	& 186.4 		& 55631	& 43	& 0.7 	& 0.7	& 119\\
J1754$-$3014 & 17:54:30.181(40) 	& $-$30:15:03(5) 		& 1.3204904144(3) 		& 4.428(19)	& 17.19 	& 89.70(7) 	& 55292	& 36	& 2.4 	& 7.6 	& 4.7\\
J1839$-$0141 & 18:39:06.9848(85)	& $-$01:41:56.0(2) 		& 0.93326558076(2) 		& 5.944(1)	& 2.53 	& 293.4		& 55467	& 13	& 2.4 	& 29 	& 2.5\\
J1848$-$1243 & 18:48:18.026(14) 	& $-$12:43:30(1) 		& 0.41438335440(2) 		& 0.4405(8)	& 3.16	& 91.96(7)	& 55595	& 9	& 0.4 	& 24		& 15\\
\tableline
\end{tabular}}
\caption{Timing solutions and derived parameters after using the wavelet denoising method for times series of eight RRATs: R.A., decl., spin period, period derivative, root-mean-square of residuals, DM, epoch of period, width of composite profile, the inferred magnetic field, and spin-down energy loss rate and characteristic age are listed. Note. The width is calculated at 50\% of the peak intensity ($\overline{W50}$). The magnetic field is at the pulsar surface and assumes alignment between spin and magnetic axis (\(B = 3.2 \times 10^{19} \sqrt{P\dot{P}}\)). The pulsar spin-down luminosity is calculated by \(\dot{E} = -4 \times 10^{46} \dot{P}/P^3\). \label{tab3}}
\end{table*}

%

\begin{table*}[!htbp]
\centering
\resizebox{\textwidth}{!}{%
\hspace{-3cm}
\begin{tabular}{lccccccccccc}
\tableline\tableline
PSR Name & R.A. & Decl. & $P$ & $\dot{P}$ & RMS & DM & Epoch & Width & $B$ & $\dot{E}$ & $\tau$\\
 & (J2000) & (J2000) & (s) & ($10^{-15}$ s s$^{-1}$) & (ms) & (pc cm$^{-3}$) & (MJD) & (ms) & ($10^{12}$ G) & ($10^{31}$ergs s$^{-1}$) & (Myr)\\
\tableline
J0735$-$6302 & 07:35:5(2)	& $-$63:02:0(4) 	& 4.862873966(3)	& 159(5) 	& 6.38 	& 19.4 		& 56212 & 24	& 28 	& 5.5 	& 0.5\\
J1048$-$5838 & 10:48:12.57(2)	& $-$58:38:18.58(15) 	& 1.231304776631(3) 	& 12.19369(7) 	& 2.22 	& 70.7(9) 	& 53510	& 7   & 3.9 	& 26 	& 1.6\\
J1226$-$3223 & 12:26:45.9(4)	& $-$32:23:14(5) 	& 6.1930029826(6) 	& 7.68(11)   	& 7.55 	& 36.7 		& 56114	& 57	& 6.0 	& 0.08 	& 20\\
J1623$-$0841 & 16:23:42.711(10) & $-$08:41:36.4(5) 	& 0.503014992514(6) 	& 1.9582(6)	& 0.76 	& 60.433(16) 	& 55079	& 13	& 1.0 	& 61 	& 4.1\\
J1739$-$2521 & 17:39:32.83(10)	& $-$25:21:2(2) 	& 1.8184611641(2) 	& 0.29(3)	& 4.85 	& 186.4 	& 55631	& 68	& 0.7 	& 0.7 	& 99\\
J1754$-$3014 & 17:54:30.08(5) 	& $-$30:14:42(6) 	& 1.3204902915(3) 	& 4.424(12)	& 3.61 	& 99.38(10) 	& 55292	& 62	& 2.4 	& 7.6 	& 4.7\\
J1839$-$0141 & 18:39:07.03(3) 	& $-$01:41:56.0(9) 	& 0.93326564072(6) 	& 5.943(3)	& 1.80 	& 293.2(6) 	& 55467	& 17	& 2.4 	& 29 	& 2.5\\
J1848$-$1243 & 18:48:17.980(8) 	& $-$12:43:26.6(5) 	& 0.41438334869(2) 	& 0.440(2)	& 1.36	& 88.0		& 55595	& 9	& 0.4 	& 24	& 15\\
\tableline
\end{tabular}
}
\caption{ Timing solutions and derived parameters for eight RRATs based on the conventional boxcar approach. Note. See Table \ref{tab3} for a description of columns.  \label{tab4}}
\end{table*}

\subsection{Pulse Profiles} 
\label{subsec:Pulse profiles}

\begin{figure}[!ht]
\centering
\includegraphics[trim=80 0 0 0,width=17cm,height=3.5cm]{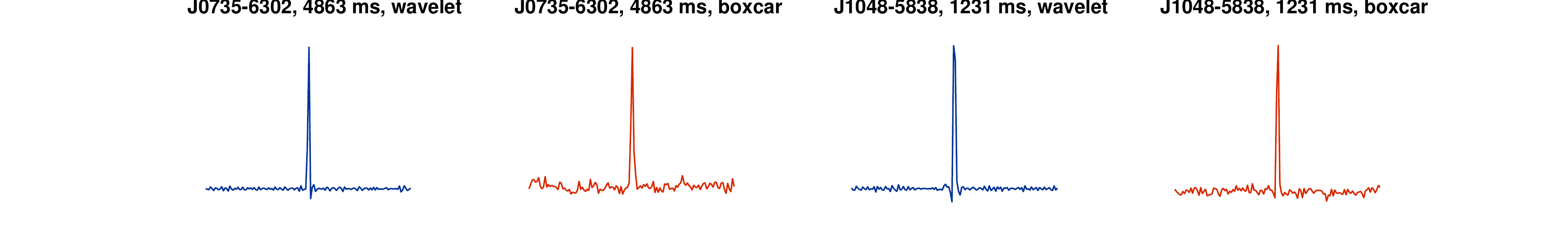}\\
\vspace{-0.3cm}
\includegraphics[trim=80 0 0 0,width=17cm,height=3.5cm]{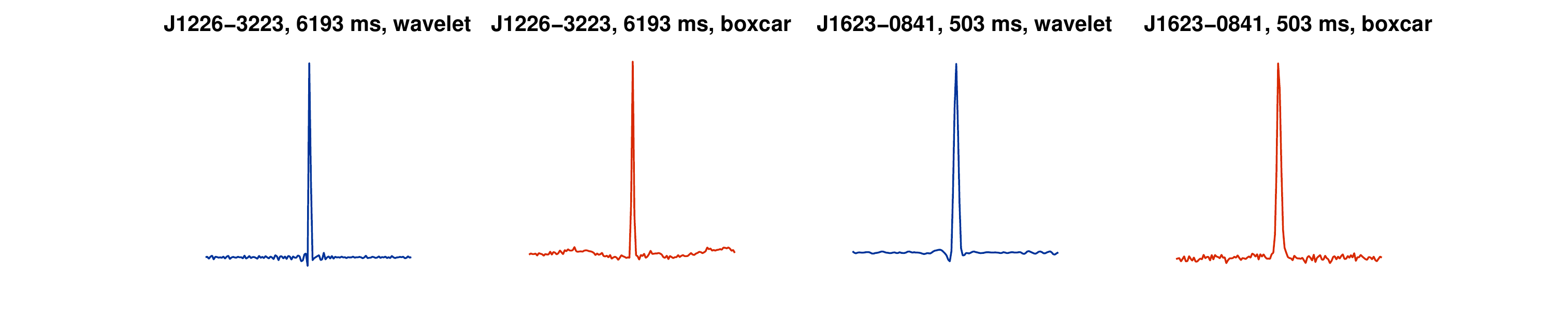}\\
\vspace{-0.3cm}
\includegraphics[trim=80 0 0 0,width=17cm,height=3.5cm]{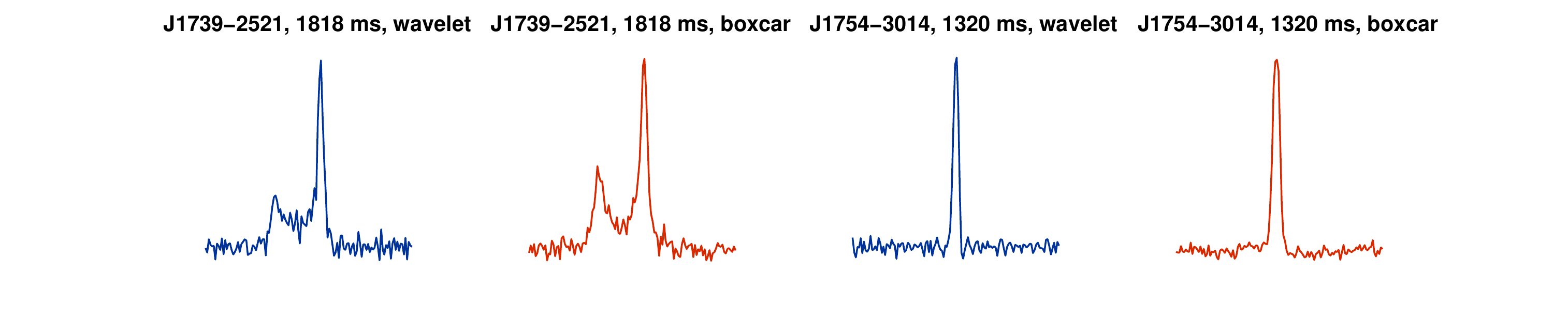}\\
\vspace{-0.3cm}
\includegraphics[trim=80 0 0 0,width=17cm,height=3.5cm]{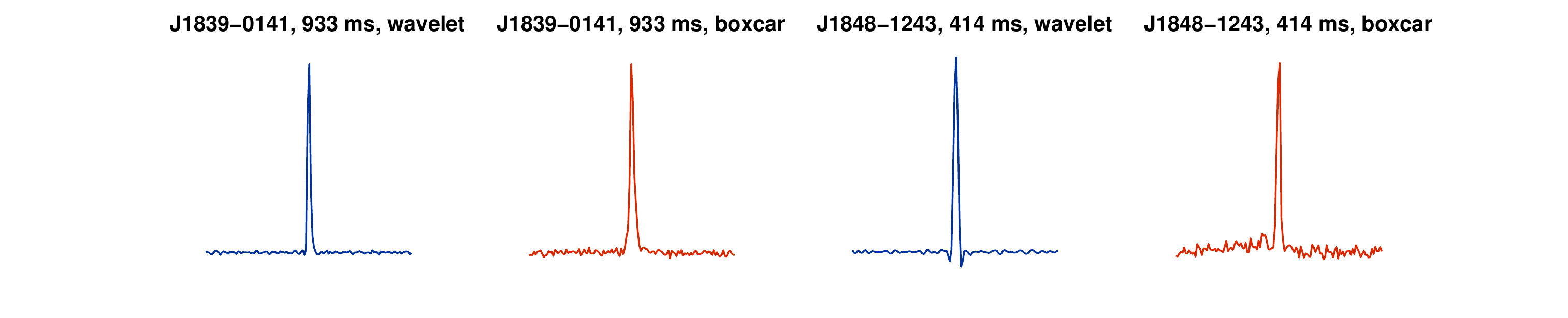}\\
\vspace{-0.5cm}
	\caption{Comparison of composite pulse profiles of eight RRATs generated from wavelet denoised data and boxcar denoised data. The wavelet denoised pulse profiles of eight RRATs are based on the denoising results in Table \ref{tab2}. Each profile is a sum of all detected individual single pulses. The spin period of each RRAT is listed above each profile. The pulse width of the profile is specified in Table \ref{tab3}. Note that after being wavelet denoised, a profile may have negative lobes on both sides. These is due to the selected shape of the mother wavelet. The composite pulse profiles which are generated from wavelet denoised data have been processed by a high-pass filter to remove the negative lobes. }
	\label{fig:pulse_profiles}
\end{figure}

We generate the composite pulse profiles by summing all detected individual single pulses, assuming the timing model was derived as shown in Section~\ref{subsec:Timing solution}. After being processed by wavelet denoising, there are negative lobes on both sides of the pulse profile. These negative lobes are not an intrinsic feature of the pulse profile. They are due to the choice of the mother wavelet shown in Figure \ref{fig:db5}. Here, we use a high-pass filter to remove these lobes. For instance, we use a high-pass filter with a cut-off frequency of 34 Hz to remove the negative fluctuations in the pulse profile of J1839$-$0141. Figure \ref{fig:pulse_profiles} shows the composite pulse profiles of eight RRATs based on the wavelet denoising results in Table \ref{tab2} and the previous boxcar denoised data from \citet{cbm+17}. The width and S/N values for these RRATs vary due to different intrinsic properties. Most of the pulse profiles become narrower after wavelet denoising (refer to the columns of Table \ref{tab3}). Initial tests show that the narrowness of the wavelet reconstructed profiles is not an intrinsic property of these RRATs but due to the removal of some parts of the full spectrum. 
\vspace{2cm}

\section{Discussion} 
\label{sec:Discussion}

\subsection{S/N Distribution} 
\label{subsec:SNR Distribution}

\begin{figure}[!ht]
\vspace{-0.1cm}
\centering
\includegraphics[width=0.42\textwidth]{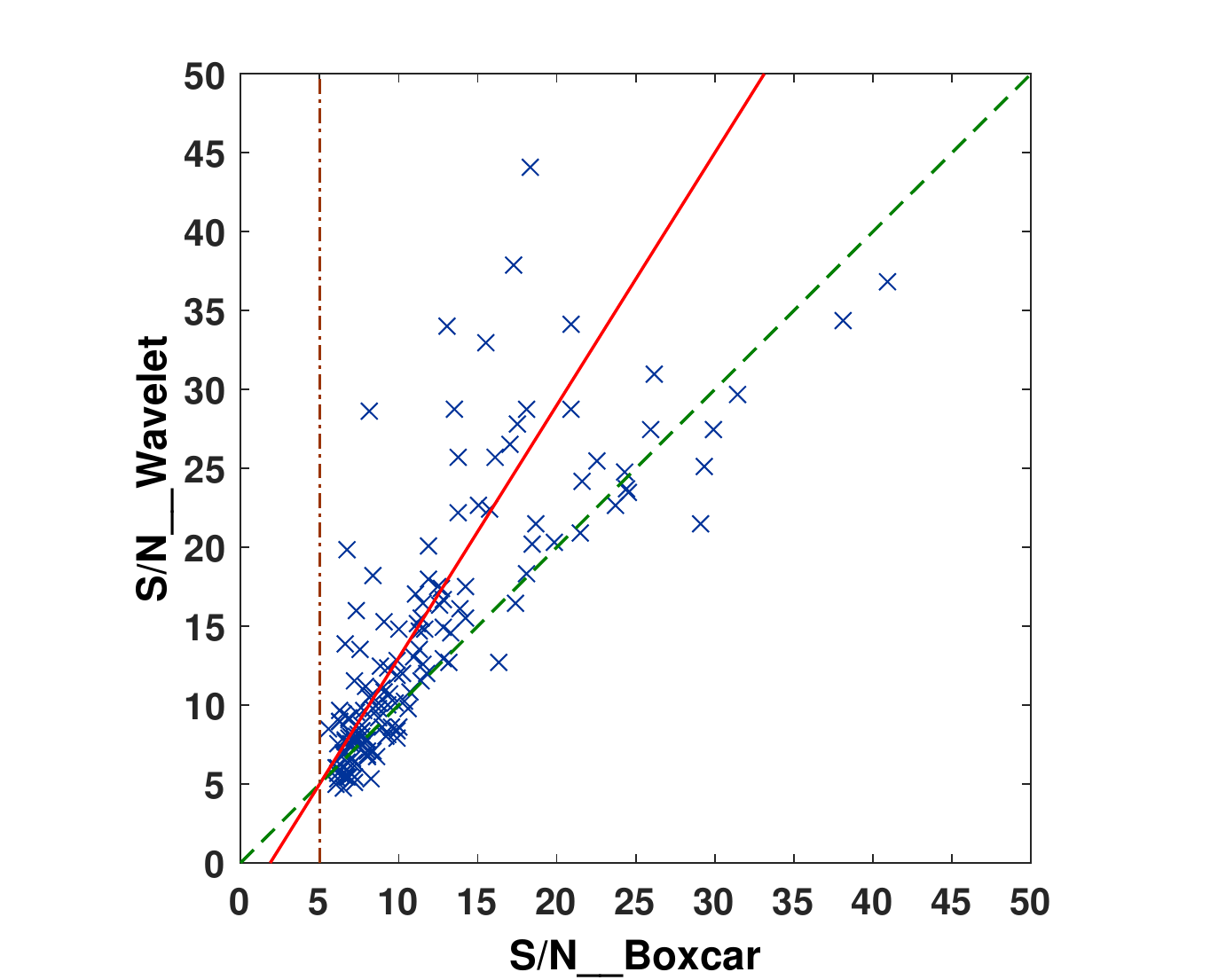}
\vspace{-0.1cm}
\caption{A comparison of single-pulse S/N calculated from J1048$-$5838's wavelet denoised signal and boxcar denoised signal. The S/N here is calculated from rebinned wavelet single-pulse profiles, which is proved to be generally larger than that calculated from boxcar single-pulse profile. The red solid line indicates a linear fit to the main distribution, the green dashed line shows where the two algorithms perform the same, and the brown dotted-dashed line shows a search threshold of 5$\sigma$. }
\label{fig:J1048_SNR}
\end{figure}

Because wavelet denoising can improve the S/N of pulses by decreasing the weights of the noise against the signal, we can detect more pulses with relatively small S/N in wavelet denoised data. Figure \ref{fig:J1048_SNR} shows a comparison of single-pulse S/N for J1048$-$5838 calculated based on the two denoising methods (boxcar and wavelet). Here, the S/N is calculated from rebinned wavelet single-pulse profiles. This is generally larger than that calculated from boxcar single-pulse profiles. However, as shown in this figure, some weak pulses could have higher S/N with boxcar denoising due to the selection effect of the threshold. The S/N threshold of the wavelet is lower than that of the boxcar due to the different sample rate and number of bins per period. Thus, some pulses detected from wavelet denoised profile (data) that are undetectable in a boxcar denoised profile are not shown in this figure. The red solid line indicates a linear fit to the main distribution, the green dashed line shows where the two algorithms perform the same, and the brown dotted-dashed line shows a search threshold of 5$\sigma$. From the distribution and the linear fitting result, we can see that the wavelet method successfully increased most of the single-pulse S/N values, especially for those originally larger than seven and less than 20. 

\subsection{Timing Solution Improvement} 
\label{subsec:Timing Solution Improvement}

In this work, there are three factors that may impact wavelet denoising's ability to improve RRAT timing solutions: the first is more pulses detected in epochs that already had some detections; the second is pulses detected in epochs that had zero detections before; and the third is sharper pulse profiles generated from wavelet denoised data. We introduce a cross-test to identify which of these three factors is responsible for the decrease in pulsar parameter errors using data on J1048$-$5838. We first fit the pulses from the wavelet denoised data that have counterparts in the boxcar results (labeled as group A, which contains the fewest number of TOAs). Then we compare this timing solution with the timing solution obtained from the boxcar denoised data (labeled as group B) and wavelet denoised data (labeled as group C, which contains the largest number of TOAs). According to Table \ref{tab5}, the average error on parameters of group A is the largest of the three, while the average error on parameters of group C is the smallest. Such differences in the timing solution errors are caused by the different number of TOAs. Here, the improvement of timing solution is largely due to the increased number of TOAs, as the number of pulses detected in epochs which had zero detections before for J1048$-$5838 data is a small fraction. The improvement of timing solution for J1048$-$5838 is largely due to the first factor.

One interesting phenomenon is that, for J1048$-$5838, the errors on R.A., decl., and $\dot{P}$ in the wavelet timing solution are all reduced compared with the timing solution of boxcar results, while the period error becomes larger. According to the pulse amplitude distribution of J1048$-$5838 (see Figure 5 in \citet{cbm+17} and Figure 8 in this paper), most newly detected pulses have relatively low S/Ns, so that the average S/N is generally lower in wavelet denoised data than that in boxcar denoised data. Because $\sigma _{TOA}\propto \rm1/(S/N)$ \citep{lk05}, lower S/N increases TOA errors. This results in a larger error on the fitted period, but lower errors on the other fitted parameters. We perform a simulation to show that this timing analysis is reasonable. The details of the simulation are given in the following paragraph.

\begin{table*}[!t]
\hspace{-2cm}
\centering
\resizebox{\textwidth}{!}{%
\begin{tabular}{lcccccc}
\tableline\tableline
 & R.A. Error & Decl. Error & $P$ Error & $\dot{P}$ Error & Number of TOAs & RMS\\
 & (J2000) & (J2000) & ($10^{-12}$ s) & ($10^{-20}$ s s$^{-1}$) & & (ms)\\
\tableline
Cross test (A) & 0.0194 & 0.156 & 5.09 & 7.06 & 183 & 2.54\\
Boxcar (B) 	& 0.0174 & 0.157 & 3.37 & 7.21 & 207 & 2.33\\
Wavelet (C)	& 0.0137 & 0.102 & 4.01 & 6.67 & 316 & 2.47\\
\tableline
\end{tabular}}
\caption{Comparison of errors in timing parameters of J1048$-$5838 calculated from the boxcar denoised data, cross-test data and wavelet denoised data. Note. Here we take a cross-test using pulses detected in both boxcar and wavelet methods. The result shows that the improvement to timing solutions using the wavelet denoising method is mainly due to the increased number of pulses detected.\label{tab5}}
\end{table*}

In the simulation, we first create two sets of fake RRAT data with the same parameters (sample time, period, S/N, and number of TOAs). In set I, we randomly delete several TOAs ($17.24\%$). For set II, we include all TOAs and set the S/N of the TOAs detected in set I to a lower value. Then we fit these two sets of data to a pulsar timing model and find that the error of the period in set II is larger. This simulation shows that even with a larger number of TOAs, the decrease in the S/N of the TOAs may still result in a larger error on the fitted period.  

The second factor (pulses detected in epochs that had zero detections before) obviously improves the timing solution for J0735$-$6302. Five epochs were detected and used for fitting in the conventional approach, while six are detected and used for the wavelet timing solution (refer to the last two columns of Table \ref{tab2}). Because five epochs were inadequate for precise fitting, the previous fitting errors are larger. In this paper, the errors on the timing parameters are dramatically decreased (see Table \ref{tab3}).

\section{Conclusions and Future Work} 
\label{sec:Conclusion and Future Work}

In this paper, we applied a wavelet denoising using selective wavelet reconstruction and wavelet shrinkage to de-dispersed radio observation data for eight RRATs. This approach suppresses the noise in the data successfully. More TOAs are able to be measured using single-pulse searching of wavelet denoised data compared with the traditional denoising approach. More precise timing solutions are obtained based on wavelet denoised RRATs data, largely due to a larger number of TOAs and epochs available for use in fitting a timing solution. On the other hand, comparing with the boxcar approach, the wavelet approach can not perform better on accurately reconstructing the pulse profiles. For further research, we will compare the speed of the wavelet approach with the boxcar approach in the same runtime environment. Then we can extrapolate using the wavelet approach for single-pulse searching and fast radio burst (FRB) searching and apply wavelet denoising to other astronomical data.

\section*{Acknowledgments} 

The work is supported by NSF Award 1458952. We would like to thank West Virginia University for its financial support for the Green Bank Telescope (GBT), which provided some of observation data in this paper. The Green Bank Observatory is a facility of the National Science Foundation operated under cooperative agreement by Associated Universities, Inc. The Parkes radio telescope is part of the Australia Telescope National Facility funded by the Australian Government for operation as a National Facility managed by CSIRO. We would also like to thank the editors and the reviewer for their comments and suggestions that improved the paper.

\facility{Green Bank Telescope, Parkes Telescope}.
\software{Matlab \citep{Matlab},
               TEMPO2 \citep{hem06}}

\bibliographystyle{aasjournal}
\bibliography{wavelet_denosie,modrefs,psrrefs}


\end{document}